\documentclass[%
  aps,
  prd,
  onecolumn,
  nofootinbib,
  preprintnumbers,
  superscriptaddress,
  floatfix
]{revtex4-2}
\usepackage[utf8]{inputenc}
\usepackage[T1]{fontenc}
\usepackage{amsmath}
\usepackage{amsfonts}
\usepackage{amssymb}
\usepackage{graphicx}
\usepackage{wrapfig} 
\usepackage{epstopdf}
\usepackage{xcolor}
\usepackage{float}
\graphicspath{{figuras/}}

\usepackage[breaklinks=true]{hyperref}
\usepackage{setspace}
\usepackage{placeins}

\definecolor{darkraspberry}{rgb}{0.53,0.15,0.34}

\newcommand{\Onod}{\Omega_{\mathrm{nod}}}
\newcommand{\Oper}{\Omega_{\mathrm{per}}}
\newcommand{\risco}{r_{\mathrm{ISCO}}}
\newcommand{\rosco}{r_{\mathrm{OSCO}}}
\newcommand{\Leff}{\Lambda_{\mathrm{eff}}}

\begin{document}

\title{Orbital stability, relativistic precession, and QPO constraints
for constant-curvature black holes in $f(R)$ gravity}

\author{Kourosh Nozari}
\email[]{nozari7450@gmail.com, knozari@umz.ac.ir}
\thanks{(Corresponding author)}
\affiliation{Department of Theoretical Physics, Faculty of Sciences,
University of Mazandaran, P.~O.~Box 47416-95447, Babolsar, Iran}

\author{Sara Saghafi}
\email[]{saghafisara1366@gmail.com}
\affiliation{Department of Theoretical Physics, Faculty of Sciences,
University of Mazandaran, P.~O.~Box 47416-95447, Babolsar, Iran}

\author{Khadijeh Salahshour}
\email[]{k.salahshour@umz.ac.ir}
\affiliation{Department of Theoretical Physics, Faculty of Sciences,
University of Mazandaran, P.~O.~Box 47416-95447, Babolsar, Iran}

\author{Mois\'es Bravo-Gaete}
\email[]{moisesbravog@gmail.com, mbravo@ucm.cl}
\affiliation{Departamento de Matem\'aticas, F\'isica y Estad\'istica, Facultad de Ciencias B\'asicas, Universidad Cat\'olica del Maule, Casilla 617,
Talca, Chile.}

\begin{abstract}
We analyze circular-orbit stability, epicyclic resonances, relativistic
precession, and quasi-periodic-oscillation (QPO) constraints in the
constant-curvature vacuum sector of metric $f(R)$ gravity. The
Schwarzschild--(anti-)de~Sitter and Kerr--(anti-)de~Sitter backgrounds
are characterized by curvature $R_0$ satisfying the algebraic trace
equation. We derive stationary-coordinate-time orbital and epicyclic
frequencies and determine stable timelike circular-orbit domains.
Positive curvature confines stable motion between innermost and
outermost stable circular orbits, which merge at a critical curvature,
and permits two resonance radii for suitable frequency ratios.
Negative curvature allows stable motion to arbitrarily large radii,
with $\Omega_\theta/\Omega_r\rightarrow1/2$, compared with unity
in the asymptotically flat limit. We determine the $3{:}2$, $2{:}1$,
and $3{:}1$ resonance branches and signed nodal precession,
recovering the Lense--Thirring limit. We also evaluate
angular-momentum-weighted rigid-flow precession and viscous alignment,
with the outermost stable orbit bounding positive-curvature flows.
A Bayesian Markov chain Monte Carlo analysis combining the simultaneous
GRO J1655--40 QPO triplet with an independent dynamical-mass measurement
yields $R_0M^2=-5.57^{+12.94}_{-16.15}\times10^{-4}$ at $68\%$
credibility, conditional on the geodesic relativistic-precession
prescription and adopted time normalization. Kerr remains allowed,
with negligible fit improvement from nonzero curvature. Doubling
the mass uncertainty broadens the curvature interval by approximately
$2.1$; removing the mass likelihood reveals an extended,
prior-dependent degeneracy. This is therefore a joint QPO and
dynamical-mass constraint, with no significant evidence for departure
from Kerr. Neutral geodesic observables depend only on $R_0$,
preserving degeneracy with Einstein gravity with an effective
cosmological constant.
\end{abstract}

\maketitle

\section{Introduction}\label{sec:intro}

In the strong-field regime, black holes (BHs) provide one of the most
powerful laboratories for testing gravity. Unlike weak-field experiments
performed within the Solar System, the spacetime surrounding compact
objects allows one to probe genuinely nonlinear gravitational phenomena,
including the existence of horizons, frame dragging, relativistic orbital
dynamics, and strong-field causal structures
\cite{Carter:1968rr,Carter:1971zc,Bardeen:1972fi}. During the last decade,
BH physics has entered an observational era, via the detection of
gravitational waves from compact binary mergers by the
LIGO Scientific and Virgo Collaborations
\cite{LIGOScientific:2016aoc,LIGOScientific:2020ibl}, the horizon-scale
imaging of supermassive BHs by the Event Horizon Telescope
\cite{EventHorizonTelescope:2019dse,EventHorizonTelescope:2022wkp}, and
increasingly precise X-ray timing observations of accreting compact
objects \cite{Remillard:2006fc,Ingram:2019mna}. Together, these
developments have transformed BHs into promising astrophysical
laboratories for testing relativistic gravity over a wide range of length
scales and dynamical regimes.

Among the various observables associated with accretion flows around BHs,
the dynamics of equatorial circular geodesics play a distinguished role
because they are entirely determined by the geometry of the underlying
spacetime. In stationary and axisymmetric configurations, circular motion
is characterized by three fundamental frequencies: the azimuthal
frequency $\Omega_{\phi}$, the radial epicyclic frequency $\Omega_{r}$,
and the vertical epicyclic frequency $\Omega_{\theta}$
\cite{Okazaki1987,Aliev:1980hz}. Small radial and vertical perturbations
of an equatorial circular orbit generate oscillatory motion governed by
the corresponding epicyclic frequencies, provided that the associated
squared frequencies remain positive. In particular, the vanishing of
$\Omega_r^2$ marks the onset of radial marginal stability and determines
the boundaries of the stable circular-orbit region, including the
innermost stable circular orbit (ISCO), one of the fundamental length
scales of relativistic accretion theory \cite{Bardeen:1972fi,Novikov1973}.
In spacetimes with a positive effective cosmological curvature, the
stable region may also possess an outer boundary, conventionally
identified as the outermost stable circular orbit (OSCO)
\cite{Stuchlik:1999qk,Stuchlik:2003dt}. Consequently, orbital frequencies
provide a unified geometric framework for describing circular-orbit
stability, relativistic precession, epicyclic resonances, and the
dynamics of accretion flows, thereby offering considerably richer
information about the underlying spacetime than the ISCO radius alone.

The intimate connection between orbital dynamics and spacetime geometry
also makes relativistic orbital frequencies a sensitive probe of
gravitational theories beyond general relativity (GR). Since the orbital
and epicyclic frequencies depend directly on the metric coefficients and
their radial derivatives, even relatively small modifications of the
background geometry may produce measurable changes in orbital stability,
epicyclic resonances, and relativistic precession
\cite{Psaltis:2008bb,Bambi:2011mj,Johannsen:2016uoh}. This interpretation
nevertheless requires some care: geodesic observables probe the spacetime
geometry directly, but they do not necessarily distinguish between
different gravitational theories that admit the same metric solution.

Quasi-periodic oscillations (QPOs) observed in the X-ray flux of BH
binaries provide a particularly interesting context in which these
fundamental frequencies may acquire astrophysical relevance. Both
low-frequency and high-frequency QPOs have been detected in accreting
stellar-mass BHs, with low-frequency QPOs conventionally classified into
types A, B, and C \cite{Remillard:2006fc,Ingram:2019mna,Casella:2005vy}.
Type-C QPOs are the most commonly observed class and display a strong
correlation with the spectral evolution of the accretion flow, whereas
high-frequency QPOs are weaker and sometimes appear in pairs with
frequency ratios close to $3 : 2$ \cite{Remillard:2006fc,Motta:2014gsa}.
Of particular relevance is GRO J1655--40, for which two high-frequency
QPOs and a simultaneous type-C QPO have been reported
\cite{Motta:2013wga}. Within the
relativistic-precession framework, such a simultaneous triplet probes
three independent combinations of the orbital and epicyclic frequencies
at a common characteristic radius and can therefore be combined with an
independent dynamical-mass measurement to constrain departures from the
Kerr geometry.
Despite extensive observational and theoretical investigation, the
physical origin of these oscillations remains unsettled. Nevertheless, a
broad family of phenomenological models relates QPO frequencies to the
characteristic frequencies of relativistic orbital motion. In the
relativistic-precession model, the observed frequencies are associated
with combinations of the fundamental geodesic frequencies, including the
periastron-precession frequency $\Oper=\Omega_\phi-\Omega_r$ and the
signed nodal-precession frequency $\Onod=\Omega_\phi-\Omega_\theta$
\cite{Motta:2013wga,Stella:1997tc,Stella:1998mq,Stella:1999sj}. The
latter originates from the loss of degeneracy between azimuthal and
vertical motion in a rotating spacetime and reduces to the standard
Lense--Thirring precession frequency in the weak-field, slow-rotation
limit. Alternatively, epicyclic-resonance models associate twin
high-frequency QPOs with nonlinear interactions between the radial and
vertical epicyclic modes; the most commonly studied parametric-resonance
condition is $\Omega_\theta/\Omega_r=3/2$, whereas the ratios
$\Omega_\theta/\Omega_r=2$ and $3$ may arise in forced-resonance
scenarios \cite{Abramowicz:2001bi,Kluzniak:2001ar,Rebusco:2004ba,Torok:2005ut}.

The use of QPO observations to probe departures from the Kerr
geometry has an established history. Early investigations considered
quasi-Kerr and rotating braneworld spacetimes
\cite{Johannsen2011QPO,Aliev2013QPO}.
Bambi~\cite{Bambi2012QPO} applied high-frequency resonance models
to non-Kerr geometries and compared their constraints with
continuum-fitting measurements. Subsequently,
Bambi~\cite{Bambi2015RPM} used the simultaneous QPO triplet
of GRO J1655--40 within the relativistic-precession model,
incorporating independent mass information to constrain
departures from Kerr. Related studies investigated
Einstein--dilaton--Gauss--Bonnet gravity
\cite{Maselli2015QPO}, Kerr--MOG black holes
\cite{Kolos2020QPO}, and Kerr black-hole mimickers
\cite{Jiang2021QPO}.

Further applications have examined noncommutative-inspired
geometries \cite{Rayimbaev2022NCQPO}, charged black holes
in Einstein--Maxwell--scalar theory
\cite{Rayimbaev2022EMSQPO}, regular Bardeen geometries
in a proposed four-dimensional Einstein--Gauss--Bonnet
setting \cite{Rayimbaev2022BardeenQPO}, hairy Horndeski
black holes \cite{Rayimbaev2023HorndeskiQPO},
and effective rotating self-dual black holes
\cite{Liu2023SelfDualQPO}. Comparative analyses have also
tested broader sets of alternative rotating spacetimes
\cite{Shahzadi2024QPO}. Environmental and nongeodesic
effects have been explored through dark-matter shells
\cite{Stuchlik2022DMQPO} and charged-particle motion
around magnetized, noncommutative-inspired black holes
\cite{Shermatov2026NCQPO}. Recent studies additionally
consider charged Kalb--Ramond black holes
\cite{Jumaniyozov2025KRQPO} and GUP-modified
Schwarzschild geometries \cite{Hoshimov2025GUPQPO}.
These investigations establish the broader context of
the present work; we do not introduce QPO-based tests
of black-hole geometry as a new observational strategy.

Relativistic precession is not restricted to individual test-particle
trajectories. If the angular-momentum axis of a geometrically thick
inner accretion flow is misaligned with the BH spin, frame-dragging
torques may cause an extended portion of the flow to undergo
approximately coherent precession. In this scenario, the global
precession frequency is obtained as an angular-momentum-weighted average
of the local nodal-precession frequency across the precessing region
\cite{Ingram:2009vm,Ingram:2011km}. Global Lense--Thirring precession has
been proposed as a possible origin of type-C QPOs and establishes a
connection between local geodesic properties and the collective dynamics
of accretion flows. Such a description, however, requires the
warp-communication timescale to be shorter than the precession timescale,
and its quantitative realization depends on the radial extent, thickness,
surface-density profile, and dissipative properties of the flow
\cite{Fragile:2007dk,Liska:2018ayk,Bollimpalli:2023loz}. Dissipative
processes may also drive the angular momentum of the flow toward
alignment with the BH spin, introducing an additional timescale that is
sensitive to the structure of the underlying spacetime
\cite{Bate:2000cn}.

Among the numerous extensions of Einstein gravity proposed over the last
decades, metric $f(R)$ gravity, where $R$ is the scalar curvature, occupies a prominent position because it
represents the simplest purely geometric generalization of the
Einstein--Hilbert action. By promoting the Ricci scalar in the
gravitational Lagrangian to a nonlinear function, these theories provide
a unified framework for addressing a wide range of problems, including
the inflationary epoch, the late-time accelerated expansion of the
Universe, and possible effective corrections arising from quantum
gravity, while reducing continuously to GR for the appropriate choice of
the function $f$ (see, e.g., Refs.
\cite{Sotiriou:2008rp,DeFelice:2010aj,Nojiri:2010wj,Capozziello:2011et}).
An especially tractable class of solutions arises in the vacuum sector
with constant scalar curvature, $R=R_0$, in which the field equations
take the Einstein-space form and $R_0/4$ plays the role of an effective
cosmological constant. The corresponding neutral static and rotating BH
geometries are locally described by Schwarzschild--(anti-)de~Sitter and
Kerr--(anti-)de~Sitter spacetimes, respectively. The underlying $f(R)$
model remains relevant because its algebraic trace equation selects the
admissible constant-curvature branches and determines their viability.
However, once a particular value of $R_0$ has been fixed, neutral
test-particle observables depend only on the resulting geometry. Thus,
different $f(R)$ models admitting the same constant-curvature branch are
degenerate at the level of neutral geodesic motion. 

Previous investigations of constant-curvature BHs in $f(R)$ gravity have
addressed several aspects of their geometry and astrophysical
phenomenology. Static and rotating solutions, including charged
generalizations, have been constructed, and their horizon structures and
thermodynamic properties analyzed
\cite{Cembranos:2011sr,Moon:2011hq,Hendi:2011hxq}. Thin accretion disks
and their radiative properties have been studied in static $f(R)$
backgrounds \cite{Pun:2008ae}, while the existence and stability of
circular orbits and the corresponding Page--Thorne spectra have been
considered for Schwarzschild- and Kerr-like constant-curvature solutions
\cite{Perez:2012bx}. Hydrodynamic equilibrium configurations, von Zeipel
surfaces, and convective stability have likewise been examined for
perfect fluids orbiting static $f(R)$ BHs \cite{Alipour:2015}. The
orbital structure of Schwarzschild--de~Sitter and Kerr--de~Sitter
spacetimes has been thoroughly analyzed in the context of GR with a
cosmological constant \cite{Stuchlik:1999qk,Stuchlik:2003dt}, providing
essential background for the present work.

QPO constraints have also been investigated in geometries containing
a cosmological-constant parameter. Rink, Caiazzo, and
Heyl~\cite{Rink2021QPO} applied the relativistic-precession model
in Kerr--Newman--de Sitter spacetime to observations of
GRO J1655--40 and XTE J1550--564, including simultaneously
measured QPO pairs. Hesamolhokama et al.~\cite{Hesamolhokama2024QPO}
performed Bayesian relativistic-precession analyses of Kerr,
standard Kerr--de Sitter, and extended Kerr--de Sitter geometries,
including the simultaneous GRO J1655--40 triplet, and examined
the degeneracy of the cosmological-constant parameter with
the other fitted quantities. These studies provide direct
precedents for the observational application considered here.

The present work builds on these earlier QPO applications and
on the established circular-orbit structure of
Schwarzschild--(anti-)de Sitter and Kerr--(anti-)de Sitter
spacetimes. Its contribution is to connect the stability boundaries
to the organization of epicyclic-resonance branches, to examine how
local nodal precession enters an angular-momentum-weighted
rigid-flow model, and to quantify the information supplied by a
simultaneous QPO triplet when combined with an independent mass
measurement. In particular, we track the occurrence and merger of
inner and outer resonance branches in the positive-curvature
domain and investigate the relation between local precession
reversal, global torque cancellation, and the limitations of the
alignment-time prescription in the negative-curvature domain.

The observational analysis complements these geometric results by
separating sensitivity to the curvature-prior width from sensitivity
to the external dynamical-mass information. It establishes a
conditional constraint on the effective curvature within the
adopted relativistic-precession prescription, rather than a
model-independent test of the function $f(R)$. The background
solutions themselves are not new, and their neutral geodesic
observables are shared by Einstein gravity with the same effective
cosmological constant. These properties raise
four closely related questions: (i)~How do the sign and magnitude of
$R_0$ determine the topology of the stable-orbit region? (ii)~Under which
conditions do epicyclic resonances possess multiple radial branches?
(iii)~How are the local nodal-precession frequency, the global
flow-precession frequency, and the viscous alignment timescale modified
by the constant background curvature? (iv)~To what extent can the
simultaneous QPO triplet of GRO J1655--40, together with its independent
dynamical-mass measurement, constrain the dimensionless constant
curvature $R_0M^2$ within the relativistic-precession model?

In this work, we address these questions for neutral massive
particles orbiting Schwarzschild--(anti-)de~Sitter and
Kerr--(anti-)de~Sitter BHs belonging to the constant-curvature
vacuum sector of metric $f(R)$ gravity. In addition to deriving the
circular-orbit stability domains, resonance branches, and local and
global precession frequencies, we confront the rotating solution
with the simultaneous QPO triplet of GRO J1655--40. Using a Bayesian
Markov chain Monte Carlo analysis based on the joint
QPO-plus-dynamical-mass posterior, we obtain
\begin{equation}
 R_0M^2
 =
 -5.57^{+12.94}_{-16.15}\times10^{-4}
 \qquad (68\%\ {\rm credible}).
\end{equation}
The Kerr value $R_0M^2=0$ lies within this interval, and its
profile-likelihood difference from the extended-model optimum is
only $\Delta\chi^2\simeq0.106$. The inferred interval is insensitive
to the tested curvature-prior width but depends appreciably on the
external dynamical-mass information: doubling the mass uncertainty
increases the width of the curvature interval by approximately a
factor of $2.1$, whereas removing the mass likelihood exposes an
extended, prior-dependent degeneracy. The primary result must
therefore be interpreted as a joint QPO and dynamical-mass
constraint on the effective constant curvature, rather than as an
independent QPO-only measurement or evidence for a nonzero value. Throughout this work, we use geometrized units, $G=c=1$, and adopt the
metric signature $(-,+,+,+)$. Analytical expressions are written in
terms of the dimensional quantities $r$, $a$, $R_0$, and $\Omega_i$,
whereas numerical figures and tables use the dimensionless variables
\begin{equation}
 x\equiv\frac{r}{M},
 \qquad
 a_\ast\equiv\frac{a}{M},
 \qquad
 \lambda\equiv R_0M^2,
 \qquad
 \bar{\Omega}_i\equiv M\Omega_i.
 \label{eq:dimensionless-conventions}
\end{equation}
Numerical times are correspondingly expressed in units of $M$.
The remainder of the paper is organized as follows. In
Sec.~\ref{sec:fR} we review the constant-curvature vacuum sector of
metric $f(R)$ gravity. In Sec.~\ref{sec:geodesics} we present the
general formalism for equatorial circular geodesics and their linear
perturbations. Section~\ref{sec:static} derives the orbital and
epicyclic frequencies and the complete classification of stable circular
orbits in the static background, and Sec.~\ref{sec:kerr} extends the
analysis to the rotating case. In Sec.~\ref{sec:resonances} we analyze
the epicyclic-frequency ratios and the corresponding resonance branches.
In Sec.~\ref{sec:precession} we investigate nodal precession, global
rigid precession, and the viscous alignment timescale. In Sec.~\ref{sec:qpo_constraints}, we confront the rotating constant-curvature geometry
with the simultaneous QPO triplet of GRO J1655--40 and an independent
dynamical-mass measurement, constrain $R_0M^2$ through an MCMC
analysis, examine its sensitivity to the curvature prior and the
external mass likelihood, and verify that the inferred QPO orbits
occupy the exterior stable branch. Finally,
Sec.~\ref{sec:conclusions} summarizes our main results and discusses
their physical interpretation and limitations.

\section{Constant-curvature vacuum sector of metric $f(R)$ gravity}
\label{sec:fR}

Metric $f(R)$ gravity provides a purely geometric extension of GR in
which the Einstein--Hilbert Lagrangian is supplemented by a nonlinear
function of the Ricci scalar
\cite{Sotiriou:2008rp,DeFelice:2010aj,Nojiri:2010wj}. To avoid ambiguity
between conventions, we write $F(R)=R+f(R)$ and consider the
four-dimensional vacuum action
\begin{equation}\label{action-fR}
S[g_{\mu\nu}]
=\frac{1}{16\pi G}\int d^4x\,\sqrt{-g}\,F(R)
=\frac{1}{16\pi G}\int d^4x\,\sqrt{-g}\,\bigl[R+f(R)\bigr],
\end{equation}
where $g$ is the determinant of the metric tensor $g_{\mu\nu}$
($\mu,\nu=0,1,2,3$) and $R$ is the scalar curvature. Varying the action
\eqref{action-fR} with respect to the metric gives the vacuum field equations
\begin{equation}\label{eq:eom1}
F'(R)R_{\mu\nu}-\frac{1}{2}F(R)g_{\mu\nu}
+\bigl(g_{\mu\nu}\Box-\nabla_\mu\nabla_\nu\bigr)F'(R)=0,
\end{equation}
where a prime denotes differentiation with respect to $R$. Taking the
trace of Eq.~\eqref{eq:eom1} yields
\begin{equation}\label{trace-equation-fR}
F'(R)R-2F(R)+3\,\Box F'(R)=0.
\end{equation}

In the present work, we restrict the analysis to vacuum solutions with
constant scalar curvature, $R=R_0=\mathrm{constant}$. In this sector,
$\Box F'(R_0)=0$ and the trace equation reduces to the algebraic
condition
\begin{equation}\label{constant-curvature-condition}
R_0F'(R_0)-2F(R_0)=0
\qquad\Longleftrightarrow\qquad
R_0\bigl[f'(R_0)-1\bigr]-2f(R_0)=0 ,
\end{equation}
so that, provided $f'(R_0)\neq 1$,
\begin{equation}\label{R0-condition}
R_0=\frac{2f(R_0)}{f'(R_0)-1}.
\end{equation}
For any nondegenerate branch satisfying $F'(R_0)=1+f'(R_0)\neq 0$, the
field equations \eqref{eq:eom1} reduce, upon using
Eq.~\eqref{constant-curvature-condition}, to the Einstein-space form
\begin{equation}\label{effective-cosmological-constant}
R_{\mu\nu}=\frac{R_0}{4}\,g_{\mu\nu}\equiv\Leff\,g_{\mu\nu},
\qquad \mbox{where}\qquad
\Leff=\frac{R_0}{4}=\frac{f(R_0)}{2\bigl[f'(R_0)-1\bigr]}.
\end{equation}
Thus, at the background level, an admissible constant-curvature vacuum
branch is described by the Einstein equations with an effective
cosmological constant. The cases $R_0>0$, $R_0=0$, and $R_0<0$
correspond, respectively, to de~Sitter, asymptotically flat, and
anti-de~Sitter branches. Once $R_0$ is fixed, neutral geodesic
observables depend only on the resulting background geometry. The
underlying $f(R)$ model nevertheless determines which values of $R_0$
solve Eq.~\eqref{constant-curvature-condition} and whether the
corresponding branch is viable. For example, $F'(R_0)>0$ ensures a positive effective gravitational coupling, while $F''(R_0)>0$ is commonly imposed to avoid the Dolgov--Kawasaki instability. The effective mass squared of the scalar degree of freedom around the constant-curvature background is 
\[
m_{\mbox{sc}}^2=\frac{F'(R_0)-R_0 F''(R_0)}{3 F''(R_0)},
\]
so the absence of a tachyonic scalaron further requires $m_{\mbox{sc}}^2>0$. These conditions are model-dependent and are logically distinct from the geodesic analysis performed below. Therefore, all orbital results derived in this work apply to any viable $f(R)$ model admitting the same constant-curvature branch, although they cannot distinguish between such models at the level of neutral test-particle motion.

\section{Test-particle motion around black holes: general formalism}
\label{sec:geodesics}

In this section we collect the standard relations for equatorial
circular geodesics and their linear radial and vertical
perturbations in stationary, axisymmetric spacetimes. The
circular-orbit analysis follows the established treatment of
Bardeen, Press, and Teukolsky~\cite{Bardeen:1972fi}
in the Kerr limit, while the epicyclic-frequency formalism
and its application to non-Kerr geometries have been developed
in earlier studies
\cite{Aliev:1980hz,Bambi2012QPO,Bambi2015RPM}.
We include the derivation to specify our notation, effective-potential
sign convention, and proper-time and coordinate-time frequency
definitions. We consider a neutral massive test particle and
neglect self-force, spin-curvature coupling, pressure, and
electromagnetic interactions. Its trajectory is therefore governed
by the geodesic equations of the background metric.

\subsection{Equatorial circular geodesics}
\label{sec:circular-geodesics}

For the stationary, axisymmetric, and circular spacetimes considered in
this work, the coordinates $(t,r,\theta,\phi)$ can be chosen such that
the line element takes the form
\begin{equation}\label{general-stationary-metric}
ds^{2}=g_{tt}\,dt^{2}+2g_{t\phi}\,dt\,d\phi+g_{\phi\phi}\,d\phi^{2}
+g_{rr}\,dr^{2}+g_{\theta\theta}\,d\theta^{2},
\end{equation}
where the metric components depend only on $r$ and $\theta$, and we
assume reflection symmetry across the equatorial plane $\theta=\pi/2$.
The stationary and axial Killing vectors $\xi_{t}=\partial_{t}$ and
$\xi_{\phi}=\partial_{\phi}$ imply that the specific energy $E$ and the
specific angular momentum $L$ are conserved along a geodesic,
\begin{equation}\label{conserved-quantities}
E=-g_{\mu\nu}\xi^{\mu}_{t}u^{\nu}=-g_{tt}u^t-g_{t\phi}u^\phi\equiv -u_{t},
\qquad
L=g_{\mu\nu}\xi^{\mu}_{\phi}u^{\nu}=g_{t\phi}u^t+g_{\phi\phi}u^\phi\equiv u_{\phi},
\end{equation}
where $u^{\mu}=dx^\mu/d\tau=(u^{t},u^{r},u^{\theta},u^{\phi})$ is the
four-velocity and $\tau$ the proper time. Defining
\begin{equation}\label{metric-discriminant}
\mathcal{D}\equiv g_{t\phi}^{2}-g_{tt}g_{\phi\phi},
\end{equation}
the inversion of Eq.~\eqref{conserved-quantities} gives
\begin{equation}\label{ut-uphi}
u^t=\frac{E\,g_{\phi\phi}+L\,g_{t\phi}}{\mathcal{D}},
\qquad
u^\phi=-\frac{E\,g_{t\phi}+L\,g_{tt}}{\mathcal{D}}.
\end{equation}
From the normalization condition for a massive particle,
$u^{\mu}u_{\mu}=-1$, we define the kinetic potential
\begin{equation}\label{kinetic-potential-def}
\mathcal{V}(r,\theta;E,L)
\equiv\frac{1}{2}\Bigl[g_{rr}(u^{r})^{2}+g_{\theta\theta}(u^{\theta})^{2}\Bigr]
=\frac{1}{2}\Bigl[-1-g_{tt}(u^{t})^{2}-g_{\phi\phi}(u^{\phi})^{2}
-2g_{t\phi}u^{t}u^{\phi}\Bigr],
\end{equation}
which, upon substituting Eq.~\eqref{ut-uphi}, becomes
\begin{equation}\label{kinetic-potential}
\mathcal{V}=\frac{1}{2}\left[-1+
\frac{E^{2}g_{\phi\phi}+2ELg_{t\phi}+L^{2}g_{tt}}{\mathcal{D}}\right].
\end{equation}
An equatorial circular orbit at $r=r_0$ and $\theta=\pi/2$ satisfies
$u^r=u^\theta=0$, together with
\begin{equation}\label{circular-orbit-conditions}
\mathcal{V}\big|_{(r_0,\pi/2)}=0,
\qquad
\partial_r\mathcal{V}\big|_{(r_0,\pi/2)}=0,
\qquad
\partial_\theta\mathcal{V}\big|_{(r_0,\pi/2)}=0.
\end{equation}
The last condition in Eq.~\eqref{circular-orbit-conditions} is automatically fulfilled by reflection symmetry. The azimuthal angular velocity with respect to the stationary coordinate
time is
\begin{equation}\label{orbital-frequency-definition}
\Omega_\phi = \frac{d\phi}{dt}=\frac{u^\phi}{u^t}.
\end{equation}
The radial geodesic equation for an equatorial circular orbit reduces to
\begin{equation}\label{circular-radial-equation}
g_{tt,r}+2\Omega_\phi\,g_{t\phi,r}+\Omega_\phi^{2}\,g_{\phi\phi,r}=0,
\end{equation}
where a comma denotes partial differentiation. Solving this quadratic
equation gives the two branches
\begin{equation}\label{orbital-frequency-general}
\Omega_\phi^{(\pm)}
=\frac{-g_{t\phi,r}\pm\sqrt{(g_{t\phi,r})^{2}-g_{tt,r}\,g_{\phi\phi,r}}}
{g_{\phi\phi,r}},
\end{equation}
corresponding to the two families of circular motion. Their
identification as prograde or retrograde depends on the sign conventions
adopted for the BH spin and the azimuthal coordinate. The existence of a
real angular velocity requires
$(g_{t\phi,r})^{2}-g_{tt,r}\,g_{\phi\phi,r}\geq0$. Once $\Omega_\phi$ is
determined, the normalization condition gives
\begin{equation}\label{ut-circular}
u^t=\frac{1}{\sqrt{-g_{tt}-2g_{t\phi}\Omega_\phi-g_{\phi\phi}\Omega_\phi^{2}}},
\end{equation}
and a timelike circular orbit exists only if the radicand is positive. The
specific energy and angular momentum of the circular orbit are then
\begin{equation}\label{EL-circular}
E=-\frac{g_{tt}+g_{t\phi}\Omega_{\phi}}
{\sqrt{-g_{tt}-2g_{t\phi}\Omega_{\phi}-g_{\phi\phi}\Omega^{2}_{\phi}}},
\qquad
L=\frac{g_{t\phi}+g_{\phi\phi}\Omega_{\phi}}
{\sqrt{-g_{tt}-2g_{t\phi}\Omega_{\phi}-g_{\phi\phi}\Omega^{2}_{\phi}}}.
\end{equation}
All metric components and their derivatives in
Eqs.~\eqref{orbital-frequency-general}--\eqref{EL-circular} are evaluated
at $r=r_0$, $\theta=\pi/2$.

\subsection{Radial and vertical epicyclic frequencies}
\label{sec:epicyclic-formalism}

The motion of a massive test particle in the neighborhood of a stable
equatorial circular orbit is characterized by the azimuthal frequency $\Omega_\phi$, which describes the
unperturbed circular motion, whereas small radial and vertical
displacements are governed by the radial and vertical epicyclic
frequencies $\Omega_r$ and $\Omega_\theta$, respectively
\cite{Okazaki1987,Aliev:1980hz}. In the geodesic approximation, these
frequencies are determined entirely by the background metric and its
derivatives.
 
We consider a trajectory neighboring the circular orbit,
\begin{equation}\label{perturbation-ansatz}
r(\tau)=r_0+\delta r(\tau),
\qquad
\theta(\tau)=\frac{\pi}{2}+\delta\theta(\tau),
\qquad
|\delta r|\ll r_0,\quad|\delta\theta|\ll1.
\end{equation}
The constants of motion are evaluated at their values on the reference circular orbit when deriving the linearized equations. More precisely, the energy excess required to sustain a finite-amplitude epicyclic oscillation is quadratic in the perturbation amplitude. Therefore, to linear order, $E$ and $L$ may consistently be replaced by their circular-orbit values. This prescription should not be interpreted as claiming that a finite-amplitude neighboring orbit has exactly the same energy as the stable circular orbit. Under this linear approximation, the perturbed motion is governed by the kinetic potential
$\mathcal{V}$ presented in Eq.~\eqref{kinetic-potential}, which by
construction satisfies
\begin{equation}\label{first-integral}
g_{rr}\,(u^{r})^{2}+g_{\theta\theta}\,(u^{\theta})^{2}
=2\,\mathcal{V}(r,\theta).
\end{equation}
Consider first a purely radial perturbation, $\delta\theta=0$, for
which Eq.~\eqref{first-integral} reduces to
$g_{rr}\,(dr/d\tau)^{2}=2\,\mathcal{V}(r,\pi/2)$. Differentiating with
respect to proper time and rearranging yields the exact
second-order equation of motion
\begin{equation}\label{radial-eom-exact}
g_{rr}\,\frac{d^{2}r}{d\tau^{2}}
+\frac{1}{2}\,g_{rr,r}\left(\frac{dr}{d\tau}\right)^{2}
=\partial_r\mathcal{V}\big|_{\theta=\pi/2}.
\end{equation}
We now insert the ansatz \eqref{perturbation-ansatz} and expand to
first order in $\delta r$. The circular-orbit conditions
\eqref{circular-orbit-conditions} guarantee that both sides of
Eq.~\eqref{radial-eom-exact} vanish on the unperturbed orbit. Moreover,
$(dr/d\tau)^{2}=\mathcal{O}(\delta r^{2})$, so the second term on the
left-hand side does not contribute at linear order, and
$\partial_r\mathcal{V}
=\partial_r^{2}\mathcal{V}|_{(r_0,\pi/2)}\,\delta r
+\mathcal{O}(\delta r^{2})$. The linearized radial equation is
therefore
\begin{equation}\label{radial-linearized}
\frac{d^{2}\delta r}{d\tau^{2}}
=\frac{1}{g_{rr}}\,
\partial_r^{2}\mathcal{V}\Big|_{(r_0,\pi/2)}\,\delta r
=-\,\omega_r^{2}\,\delta r ,
\end{equation}
which defines the {proper-time} radial epicyclic frequency
$\omega_r$. The same steps applied to a purely vertical perturbation,
$\delta r=0$, using
$g_{\theta\theta}\,(d\theta/d\tau)^{2}=2\,\mathcal{V}(r_0,\theta)$,
give
\begin{equation}\label{vertical-linearized}
\frac{d^{2}\delta\theta}{d\tau^{2}}
=\frac{1}{g_{\theta\theta}}\,
\partial_\theta^{2}\mathcal{V}\Big|_{(r_0,\pi/2)}\,\delta\theta
\equiv-\,\omega_\theta^{2}\,\delta\theta ,
\end{equation}
where reflection symmetry across the equatorial plane ensures that
$\theta=\pi/2$ is an equilibrium plane,
$\partial_\theta\mathcal{V}|_{(r_0,\pi/2)}=0$. Reflection symmetry also
implies $\partial_r\partial_\theta\mathcal{V}|_{(r_0,\pi/2)}=0$, so
that for a general perturbation the radial and vertical oscillations
decouple at linear order and Eqs.~\eqref{radial-linearized}--\eqref{vertical-linearized} hold simultaneously.
 
Finally, we convert the proper-time frequencies to frequencies measured with respect to the stationary coordinate time. Along the perturbed trajectory,  $dt/d\tau=u^t+\mathcal{O}(\delta r,
\delta\theta)$, with $u^t$ the circular-orbit value given in Eq.
\eqref{ut-circular}. Since $\delta r$ and $\delta\theta$ are already
first-order quantities, the correction contributes only at second
order, and $d/d\tau=u^t\,d/dt$ may be used with $u^t$ held constant.
The perturbations then obey two decoupled harmonic equations:
\begin{equation}\label{harmonic-equations}
\frac{d^2\delta r}{dt^2}+\Omega_r^2\,\delta r=0,
\qquad
\frac{d^2\delta\theta}{dt^2}+\Omega_\theta^2\,\delta\theta=0,
\end{equation}
with the coordinate-time epicyclic frequencies
\begin{equation}\label{frequency-original-potential}
\Omega_r^2=\frac{\omega_r^2}{(u^t)^2}
=-\frac{1}{g_{rr}\,(u^t)^2}\,
\partial_r^2\mathcal{V}\Big|_{(r_0,\pi/2)},
\qquad
\Omega_\theta^2=\frac{\omega_\theta^2}{(u^t)^2}
=-\frac{1}{g_{\theta\theta}\,(u^t)^2}\,
\partial_\theta^2\mathcal{V}\Big|_{(r_0,\pi/2)},
\end{equation}
where $u^t$, $g_{rr}$, and $g_{\theta\theta}$ are evaluated on the
unperturbed circular orbit. Since the proper-time and coordinate-time
frequencies differ by the common factor $u^t$, frequency ratios
are independent of this choice, whereas individual frequencies are not.
 
A stable circular orbit sits at a local maximum of the
kinetic potential along each direction
($\partial^2\mathcal{V}<0$, recall that $\mathcal{V}=0$ on the orbit
and $\mathcal{V}\geq0$ wherever motion is allowed), and the curvature
of $\mathcal{V}$ at the orbit plays the role of the restoring-force
constant. Radial and vertical stability require
\begin{equation}\label{stability-conditions}
\Omega_r^2>0,
\qquad
\Omega_\theta^2>0 .
\end{equation}
The condition $\Omega_r^2=0$ determines a radially marginally stable
circular orbit. Depending on the asymptotic structure of the spacetime,
this condition may define an ISCO, an OSCO, or both. Likewise,
$\Omega_\theta^2=0$ signals the onset of vertical marginal stability.
In the Newtonian limit of a spherically symmetric potential $1/r$, the three
frequencies coincide, $\Omega_\phi=\Omega_r=\Omega_\theta$, and the
orbits close. The differences among them are thus
relativistic (or, as we shall see, curvature-induced) effects, and it
is precisely these differences that source the periastron and nodal
precessions discussed in Secs.~\ref{sec:precession} and
\ref{sec:qpo_constraints}. In the following
two sections, we evaluate these quantities for the
Schwarzschild--(anti-)de~Sitter and Kerr--(anti-)de~Sitter geometries of
the constant-curvature vacuum sector.

\section{Orbits and epicyclic frequencies in the
Schwarzschild--(anti-)de~Sitter background}
\label{sec:static}

We now specialize the general formalism to the static branch. In
coordinates $(t,r,\theta,\phi)$, the constant-curvature static solution
takes the form
\begin{equation}\label{static-metric}
ds^2=-h(r)\,dt^2+\frac{dr^2}{h(r)}
+r^2\bigl(d\theta^2+\sin^2\theta\,d\phi^2\bigr),
\qquad
h(r)=1-\frac{2M}{r}-\frac{R_0}{12}\,r^2 ,
\end{equation}
which is the Schwarzschild--de~Sitter ($R_0>0$),
Schwarzschild ($R_0=0$), or Schwarzschild--anti-de~Sitter ($R_0<0$)
geometry with $\Leff=R_0/4$, and $M$ an integration constant related to the mass. In the limit $R_0\to0$ the constant $M$ coincides
with the ADM mass of the asymptotically flat solution. For $R_0\neq0$,
the spacetime is not asymptotically flat and $M$ must be interpreted as
the mass {parameter} of the solution (e.g.\ in the sense of the
quasi-local or Abbott--Deser-type constructions), a distinction that is
immaterial for the geodesic analysis below, which involves only the
metric function \eqref{static-metric}. For $R_0<0$ it is convenient to
note that $h(r)\to r^2/\ell^2$ at large radii, with effective
anti-de~Sitter curvature radius $\ell^2=-12/R_0$.
 
The horizon structure follows from the real positive roots of
$h(r)=0$, i.e.\ of the cubic
\begin{equation}\label{horizon-cubic}
\frac{R_0}{12}\,r^{3}-r+2M=0 .
\end{equation}
For $R_0>0$, Eq.~\eqref{horizon-cubic} possesses two positive roots
provided $0<R_0M^2<\frac{4}{9}$,
namely the BH horizon $r_h$ and the cosmological horizon $r_c$. For small curvature ($R_0M^2\ll1$), these reduce to
\begin{equation}\label{horizon-expansions}
r_h\simeq2M\left(1+\frac{R_0M^2}{3}\right),
\qquad
r_c\simeq\sqrt{\frac{12}{R_0}}-M ,
\end{equation}
showing that the cosmic repulsion pushes the BH horizon slightly
outward while the cosmological horizon recedes as
$R_0^{-1/2}$. The two horizons approach each other as $R_0$ grows and
merge in the extreme (Nariai) limit $R_0M^2=4/9$, at $r=3M$. Beyond
this value the solution describes a naked singularity embedded in
de~Sitter space, and we shall not consider it further.

\subsection{Circular orbits}
\label{sec:static-circular}
For equatorial motion, the conserved quantities in Eq.~\eqref{conserved-quantities} give

\begin{equation}
u^t=\frac{E}{h(r)}, \quad u^\phi=\frac{L}{r^2},\label{static-ut-uphi}
\end{equation}
and the normalization condition yields
the radial equation
\begin{equation}\label{static-radial-energy-equation}
(u^r)^2+\,\mathcal{U}(r)=E^2,
\qquad
\mathcal{U}(r)=h(r)\left(1+\frac{L^2}{r^2}\right).
\end{equation}
Here, $\mathcal{U}(r)$ is the conventional radial effective potential. It is important to note that we
reserve the symbol $\mathcal{V}$ for the kinetic potential of
Eq.~\eqref{kinetic-potential}. The two functions encode the same radial
dynamics with different normalizations and are related by
\[
\mathcal{V}=\frac{E^2-\mathcal{U}}{2h}.
\]
Thus, the two potentials describe exactly the same radial dynamics but use opposite stability conventions: a stable circular orbit is a local minimum of $\mathcal{U}$ and a local maximum of $\mathcal{V}$. In all radial derivatives, $E$ and $L$ are held fixed at their values on the reference circular orbit.

Circular orbits ($r=\mathrm{constant}$ and $u^{r}=\dot{u}^{r}=0$) are located at
the extrema of $\mathcal{U}$ and satisfy
\begin{equation}\label{static-circular-conditions}
\mathcal{U}(r)=E^2,
\qquad
\mathcal{U}'(r)=h'\left(1+\frac{L^2}{r^2}\right)-\frac{2hL^2}{r^3}=0,
\end{equation}
giving the expressions
\begin{equation}\label{static-EL}
E^2=\frac{2h^2(r)}{2h(r)-r\,h'(r)}=\frac{r \left(1-\frac{2M}{r}-\frac{R_0}{12}\,r^2\right)^2}{r-{3M}},
\qquad
L^2=\frac{r^3h'(r)}{2h(r)-r\,h'(r)}
=\frac{r^2\left(M-\dfrac{R_0}{12}r^3\right)}{r-3M},
\end{equation}
where the second equalities follow from the identity
\begin{equation}\label{photon-identity}
2h(r)-r\,h'(r)=2\left(1-\frac{3M}{r}\right),
\end{equation}
valid for the metric function \eqref{static-metric}. This equation has an immediate
geometric consequence: since $E^2$ and $L^2$ diverge as
$r\to3M$, circular geodesics become null there, and the photon
sphere remains at $r_{\rm ph}=3M$ for all values of $R_0$, as
is well known for (anti-)de~Sitter-Schwarzschild spacetimes
\cite{Stuchlik:1999qk}. The constant background curvature displaces
the horizons and the marginally stable orbits, but not the circular
photon orbit. Furthermore, Eq.~\eqref{static-EL} makes the domain of existence of circular
orbits transparent. Timelike circular orbits require $r>3M$, so that
$E^2>0$ and the denominators are positive, together with $L^2\geq0$.
For $R_0\leq0$ the latter condition holds for all $r>3M$, and the
circular-orbit family extends to arbitrarily large radii. However, its
asymptotic behavior is drastically modified with respect to
the asymptotically flat case, since
$L^2\simeq(|R_0|/12)\,r^{4}$ as $r\to+\infty$ (i.e.\
$L\simeq r^{2}/\ell$). At large radii the anti-de~Sitter background
acts as an isotropic harmonic trap of frequency $1/\ell$, a point to
which we return below. For $R_0>0$, instead, $L^2\geq0$ requires
\begin{equation}\label{static-radius}
r\leq r_s\equiv\left(\frac{12M}{R_0}\right)^{1/3},
\end{equation}
where $r_s$ is the {static radius}, at which the specific angular
momentum vanishes. A particle at $r=r_s$ can remain at rest, gravity
being exactly balanced by the cosmic repulsion, and no circular
geodesics exist beyond it.

\subsection{Orbital and epicyclic frequencies}
\label{sec:static-frequencies}

The three fundamental frequencies now follow from the general
formalism of Sec.~\ref{sec:geodesics}, and each admits a short
independent derivation that serves as a cross-check. For the
azimuthal frequency, Eq.~\eqref{orbital-frequency-general} with
$g_{t\phi}=0$, $g_{tt}=-h$, and $g_{\phi\phi}=r^2$, gives
$\Omega_\phi^2=-g_{tt,r}/g_{\phi\phi,r}=h'/(2r)$. Equivalently,
$\Omega_\phi=u^\phi/u^t=h L/(E r^{2})$ via Eq.~\eqref{static-ut-uphi}. For the vertical frequency, the second
$\theta$-derivative of the kinetic potential
\eqref{kinetic-potential} at the equatorial plane evaluates to
$\partial^2_\theta\mathcal{V}|_{\pi/2}=-L^2/r^2$, so that
Eq.~\eqref{frequency-original-potential} yields
\begin{equation}\label{static-Omega-theta-raw}
\Omega_\theta^2
=\frac{L^{2}}{r^{4}\,(u^t)^{2}}
=\frac{h^{2}L^{2}}{E^{2}r^{4}}.
\end{equation}
Using the circular-orbit expressions for $E$ and $L$, we obtain:
\[
\Omega^2_\phi=\frac{h'(r)}{2r}=\Omega^2_\theta,
\]
which is an exact consequence of spherical symmetry: a slightly tilted circular orbit is equivalent to a circular orbit in a rotated orbital plane. For the radial frequency, the one-dimensional form
\eqref{static-radial-energy-equation} gives directly
$\omega_r^2=\tfrac{1}{2}\,\mathcal{U}''(r)$ at fixed $L$, and the
conversion factor to coordinate time is
\begin{equation}\label{static-ut-factor}
(u^t)^{2}=\frac{E^{2}}{h^{2}}=\frac{r}{r-3M},
\end{equation}
in which, remarkably, all curvature terms cancel by virtue of
Eq.~\eqref{photon-identity}. Collecting these results, the
coordinate-time frequencies read
\begin{equation}\label{static-Omega-phi-theta}
\Omega_\phi^2=\Omega_\theta^2=\frac{h'(r)}{2r}
=\frac{M}{r^3}-\frac{R_0}{12},
\end{equation}
\begin{equation}\label{static-Omega-r}
\Omega_r^2
=\frac{1}{2r}\Bigl(r h(r) h''(r)-2r [h'(r)]^2+3h(r)h'(r)\Bigr)=\frac{M}{r^3}-\frac{6M^2}{r^4}
+\frac{5MR_0}{4r}-\frac{R_0}{3}.
\end{equation}

For the numerical analysis and figures, we introduce the dimensionless
variables
\begin{equation}
 x\equiv\frac{r}{M},
 \qquad
 a_\ast\equiv\frac{a}{M},
 \qquad
 \lambda\equiv R_0M^2.
 \label{eq:dimensionless-variables}
\end{equation}
In geometrized units, the orbital and epicyclic frequencies have dimensions
of inverse mass, $[\Omega_i]=M^{-1}$. We therefore report numerical
frequencies through the dimensionless combinations
$\bar{\Omega}_i\equiv M\Omega_i$. In particular, Eqs.~(39) and (40)
become
\begin{align}
 \left(M\Omega_\theta\right)^2
 &=
 \frac{1}{x^3}-\frac{\lambda}{12},
 \label{eq:dimensionless-static-vertical}
 \\
 \left(M\Omega_r\right)^2
 &=
 \frac{1}{x^3}
 -\frac{6}{x^4}
 +\frac{5\lambda}{4x}
 -\frac{\lambda}{3}.
 \label{eq:dimensionless-static-radial}
\end{align}
Consequently, whenever numerical results are displayed as functions of
$r/M$ and $R_0M^2$, the corresponding dimensionless frequency axes are
$M\Omega_r$, $M\Omega_\theta$, $M\Omega_{\rm nod}$, and $M\Omega_p$.
Dimensionless times are similarly reported in units of $M$. We retain the
dimensional notation $r$, $a$, $R_0$, and $\Omega_i$ in the analytical
expressions unless dimensionless variables are explicitly introduced.
 
Several features of
Eqs.~\eqref{static-Omega-phi-theta}--\eqref{static-Omega-r} deserve
comment: 
(i)~Setting $R_0=0$ recovers the familiar Schwarzschild results
$\Omega_\phi^2=M/r^3$ and $\Omega_r^2=(M/r^3)(1-6M/r)$, while for
$M/r\ll1$ and $|R_0|r^2\ll M/r$ both reduce to the Newtonian
Keplerian frequency. (ii)~The curvature enters $\Omega_\phi^2$
through the additive shift $-R_0/12$, whose {fractional} size
relative to the Keplerian term grows as $|R_0|r^3/(12M)$. Constant
background curvature is a large-radius effect, negligible near the
ISCO for the small values of $R_0M^2$ considered here and dominant in
the outer disk. (iii)~$\Omega_\phi^2$
vanishes precisely at the static radius \eqref{static-radius},
consistently with $L\to0$ there. (iv)~In the opposite, curvature-dominated
regime, formally $M\to0$, or $r\to+\infty$ with $R_0<0$, one finds
$\Omega_\phi=\Omega_\theta\to1/\ell$ and $\Omega_r\to2/\ell$ with
$\ell^2=12/|R_0|$. These are exactly the frequencies of an isotropic
harmonic oscillator, whose bound orbits are ellipses centered at the
origin and therefore traverse two radial oscillations per revolution.

\begin{figure*}[htbp]
\centering
\includegraphics[width=1\textwidth]{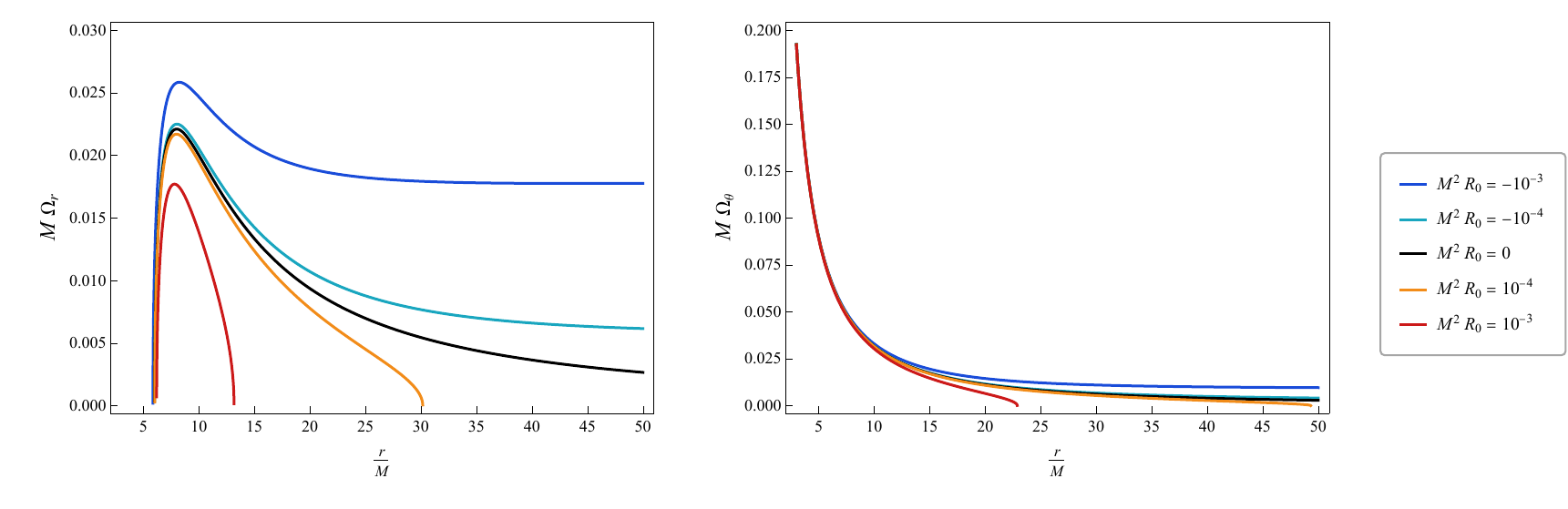}
{\caption{Dimensionless radial and vertical epicyclic frequencies,
$M\Omega_{r}$ and $M\Omega_{\theta}$, respectively, as functions of
the dimensionless radial coordinate $r/M$ in the Schwarzschild--(anti-)de
Sitter background of the constant-curvature vacuum sector of metric
$f(R)$ gravity. The left panel shows $M\Omega_{r}$ for representative
values of the dimensionless constant curvature $R_{0}M^{2}$, with the
locations of the maxima depending on $R_{0}M^{2}$. The right panel shows
$M\Omega_{\theta}$ for the same curvature values. The general-relativistic
Schwarzschild limit, $R_{0}M^{2}=0$, is included for comparison.}}
\label{fig1}
\end{figure*}

Figure \ref{fig1} illustrates that the effect of the constant curvature is predominantly a large-radius phenomenon. Near the inner stable orbit, the curves remain close to their Schwarzschild values for the small dimensionless curvatures considered here.  For positive $R_0$, 
the radial frequency vanishes at the ISCO and again at the OSCO, confining radially stable motion to a finite interval. The vertical frequency decreases outward and vanishes at the static radius $r_s=(12M/R_0)^{1/3}$, where the angular momentum of the circular orbit approaches zero. For negative $R_0$, stable circular motion extends toward arbitrarily large radii. In the curvature-dominated region, $\Omega^2_r\to {|R_0|/3}$, $\Omega^2_\phi=\Omega^2_\theta\to{|R_0|/12}$. Thus, $\Omega_\theta/\Omega_r\to {1/2}$, which differs qualitatively from the asymptotically flat Schwarzschild limit, where the ratio approaches unity. These modifications are geometric consequences of the constant background curvature and are not unique signatures of the underlying $f(R)$ theory.

\subsection{Stability domain: ISCO, OSCO, and the critical curvature}
\label{sec:static-stability}

\begin{figure*}[htbp]
\centering
\includegraphics[width=0.6\textwidth]{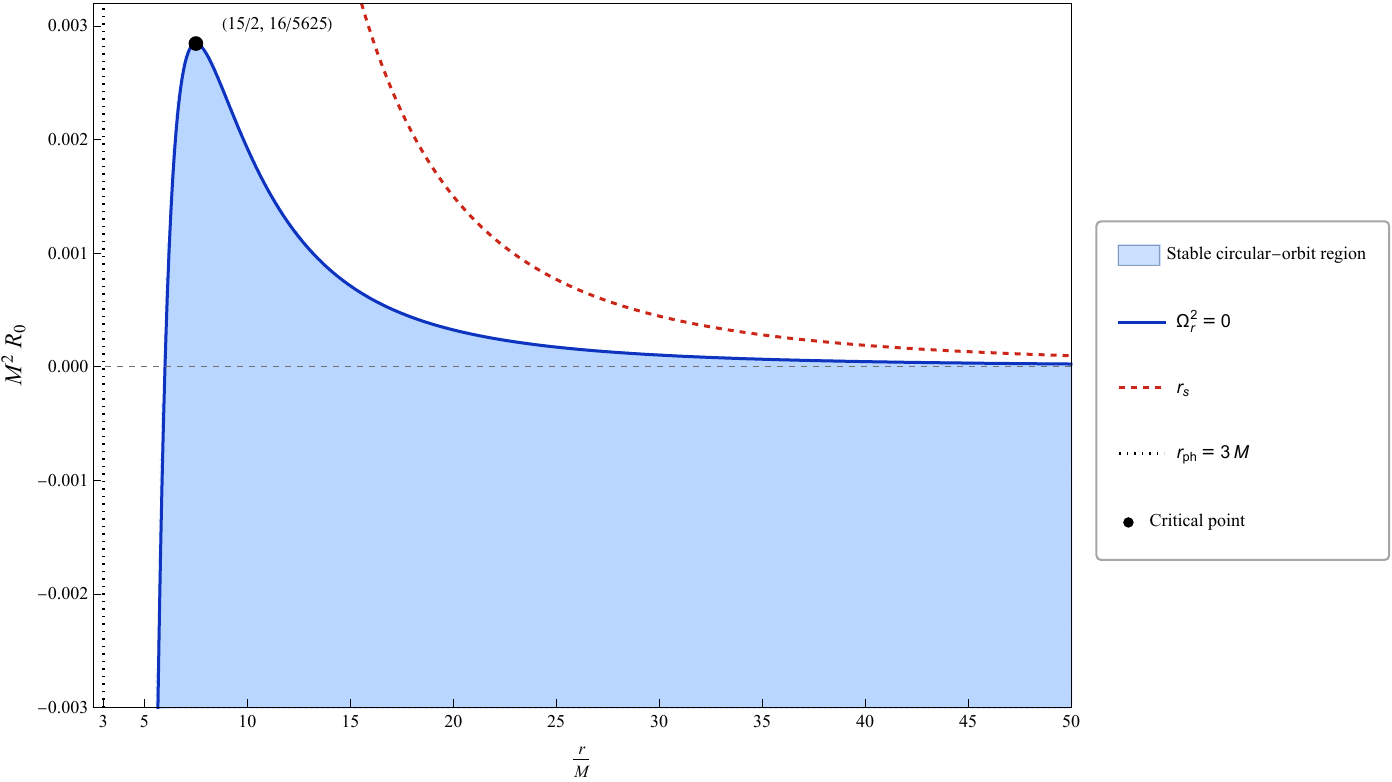}
\caption{Stability domain of equatorial circular geodesics in the
$(r/M,\,R_{0}M^{2})$ plane for the static background (shaded region).
The solid curve represents the marginal-stability condition
$(M\Omega_{r})^{2}=0$. Its left, steep branch corresponds to the ISCO,
which continues smoothly into the $R_{0}M^{2}<0$ region at radii
slightly smaller than $r/M=6$, while its descending branch for
$R_{0}M^{2}>0$ corresponds to the OSCO. The two branches merge at the
critical point
$(r/M,R_{0}M^{2})=(15/2,\,16/5625)$, above which no stable circular
orbit exists. The dashed curve denotes the dimensionless static radius,
$r_{s}/M=\left(12/(R_{0}M^{2})\right)^{1/3}$,
which forms the outer boundary of the circular-orbit domain but is not
itself part of the stable region. The dotted vertical line marks the
curvature-independent photon sphere at $r_{\rm ph}/M=3$.}
\label{fig:stability-domain}
\end{figure*}

Radial marginal stability, $\Omega_r^2=0$, is governed by the quartic
polynomial obtained from Eq.~\eqref{static-Omega-r}, or
equivalently
\begin{equation}\label{marginal-stability-curve}
R_0(r)=\frac{12M\,(r-6M)}{r^3\,(4r-15M)} ,
\end{equation}
which expresses the marginally stable radii as level curves of a single
function. The structure of the stable region follows directly.

(i)~For negative curvature ($R_0<0$), $\Omega_r^2=0$ possesses a single
positive root $\risco$, slightly below the Schwarzschild value. To first
order in $|R_0|M^2\ll1$,
\begin{equation}\label{isco-expansion}
\risco\simeq 6M\bigl(1+27\,R_0M^2\bigr).
\end{equation}
Stable circular orbits fill the entire range $r>\risco$. The stable
region is noncompact, as in the asymptotically flat case.

(ii)~For positive curvature ($0<R_0M^2<\left(R_0M^2\right)_{\rm crit},$),
the function \eqref{marginal-stability-curve} has a single maximum,
attained at $r=15M/2$, where it takes the value:
\begin{equation}\label{critical-curvature}
\left(R_0M^2\right)_{\rm crit}
=
\frac{16}{5625}.
\end{equation}
For $R_0<R_0^{\rm crit}$, the equation $\Omega_r^2=0$ has two roots in the circular-orbit domain, the ISCO
($\risco$), given by Eq.~\eqref{isco-expansion} with $R_0>0$ (the ISCO
moves {outward}), and the OSCO ($\rosco$), whose small-curvature expansion
reads
\begin{equation}\label{osco-expansion}
\rosco\simeq\left(\frac{3M}{R_0}\right)^{1/3}-\frac{3}{4}M ,
\end{equation}
consistently with the known Schwarzschild--de~Sitter result
\cite{Stuchlik:1999qk}. Stable circular orbits exist only in the compact
annulus $\risco<r<\rosco$, which shrinks as $R_0$ grows and disappears
at the critical curvature \eqref{critical-curvature}, where ISCO and
OSCO merge at $r=15M/2$. Note the hierarchy
$3M<\risco<\rosco<r_s<r_c$. In particular,
$\rosco/r_s=(3/12)^{1/3}\simeq0.63$ at leading order, so the stable
annulus always lies well inside the static radius.

(iii)~Vertical stability. Since
$\Omega_\theta^2=\Omega_\phi^2>0$ throughout the circular-orbit domain
(see Eq.~\eqref{static-Omega-phi-theta}), vertical stability imposes no additional restriction in the static case. Figure~\ref{fig:stability-domain} summarizes the stability domain in
the $(r/M,R_0M^2)$ plane.

\section{Orbits and epicyclic frequencies in the
Kerr--(anti-)de~Sitter background}
\label{sec:kerr}

The rotating member of the constant-curvature vacuum sector is the
Kerr--(anti-)de~Sitter geometry, which in Boyer--Lindquist-type
coordinates reads \cite{Stuchlik:2003dt,Cembranos:2011sr}
\begin{align}
ds^2={}&
\frac{\rho^2}{\Delta_r}\,dr^2
+\frac{\rho^2}{\Delta_\theta}\,d\theta^2
+\frac{\Delta_\theta\sin^2\theta}{\rho^2}
\left[
\frac{a\,dt}{\Xi}
-\frac{r^2+a^2}{\Xi}\,d\phi
\right]^2
-
\frac{\Delta_r}{\rho^2}
\left[
\frac{dt}{\Xi}
-\frac{a\sin^2\theta}{\Xi}\,d\phi
\right]^2.
\label{eq:kerr-ads-metric}
\end{align}
Here,
\begin{equation}\label{kerr-ds-functions}
\Delta_{r}=(r^{2}+a^{2})\left(1-\frac{R_0}{12}r^{2}\right)-2Mr,
\qquad
\Delta_{\theta}=1+\frac{R_0}{12}a^{2}\cos^{2}\theta,
\qquad
\rho^{2}=r^{2}+a^{2}\cos^{2}\theta,
\qquad
\Xi=1+\frac{R_0}{12}a^{2},
\end{equation}
where $a$ is the angular momentum per unit BH mass. For $R_0=0$, the
metric reduces to Kerr, while for $a=0$ it reduces to the static
solution~\eqref{static-metric}. In the coordinate convention adopted
in Eq.~\eqref{eq:kerr-ads-metric}, both the stationary and azimuthal
coordinate differentials are normalized by the constant factor $\Xi$,
so that the metric contains the combinations $dt/\Xi$ and
$d\phi/\Xi$. This common factor cancels from the circular-orbit
condition~\eqref{orbital-frequency-general}, leading to the angular
velocity given later in Eq.~\eqref{kerr-ds-orbital-frequency}.

An alternative stationary-time convention may be introduced through
$\hat{t}=t/\Xi$, while leaving the azimuthal coordinate unchanged. The
frequencies in the two conventions are related by
\[
\Omega_i^{(\hat{t})}
=
\Xi\,\Omega_i^{(t)},
\qquad
i\in\{\phi,r,\theta\}.
\]

Consequently, the marginal-stability radii and epicyclic-frequency
ratios are unchanged by this common constant rescaling of time.
This statement concerns a change of coordinate normalization at
fixed azimuthal coordinate; it does not establish independence
from an arbitrary observer or rotating reference frame.
For example, under
\begin{equation}
 t'=A t,\qquad
 \varphi'=\varphi-\omega_{\rm f}t,
 \qquad A>0,
\end{equation}
where $A$ and $\omega_{\rm f}$ are constants, the orbital and
epicyclic frequencies transform as
\begin{equation}
 \Omega'_\varphi=\frac{\Omega_\varphi-\omega_{\rm f}}{A},
 \qquad
 \Omega'_r=\frac{\Omega_r}{A},
 \qquad
 \Omega'_\theta=\frac{\Omega_\theta}{A}.
\end{equation}
Thus, $\Omega_\theta/\Omega_r$ is unchanged by this transformation,
whereas precession rates defined relative to the azimuthal frame
also depend on its rotation.

We use the stationary time and azimuthal coordinate of the metric
above throughout the calculation. This choice recovers the
standard Boyer--Lindquist convention in the Kerr limit and provides
a definite reference for comparing the curvature branches.
However, smooth recovery of the Kerr limit does not uniquely
select an observational clock at nonzero curvature. The de~Sitter
branch has no asymptotically flat static infinity, while the
anti-de~Sitter branch requires a choice of boundary time and
rotational frame. The coordinate convention therefore defines
the frequencies calculated here, but does not alone determine
their identification with frequencies measured by a distant
observer.

\subsection{Circular orbits and fundamental frequencies}

\begin{figure*}[htbp]
\centering
\includegraphics[width=1\textwidth]{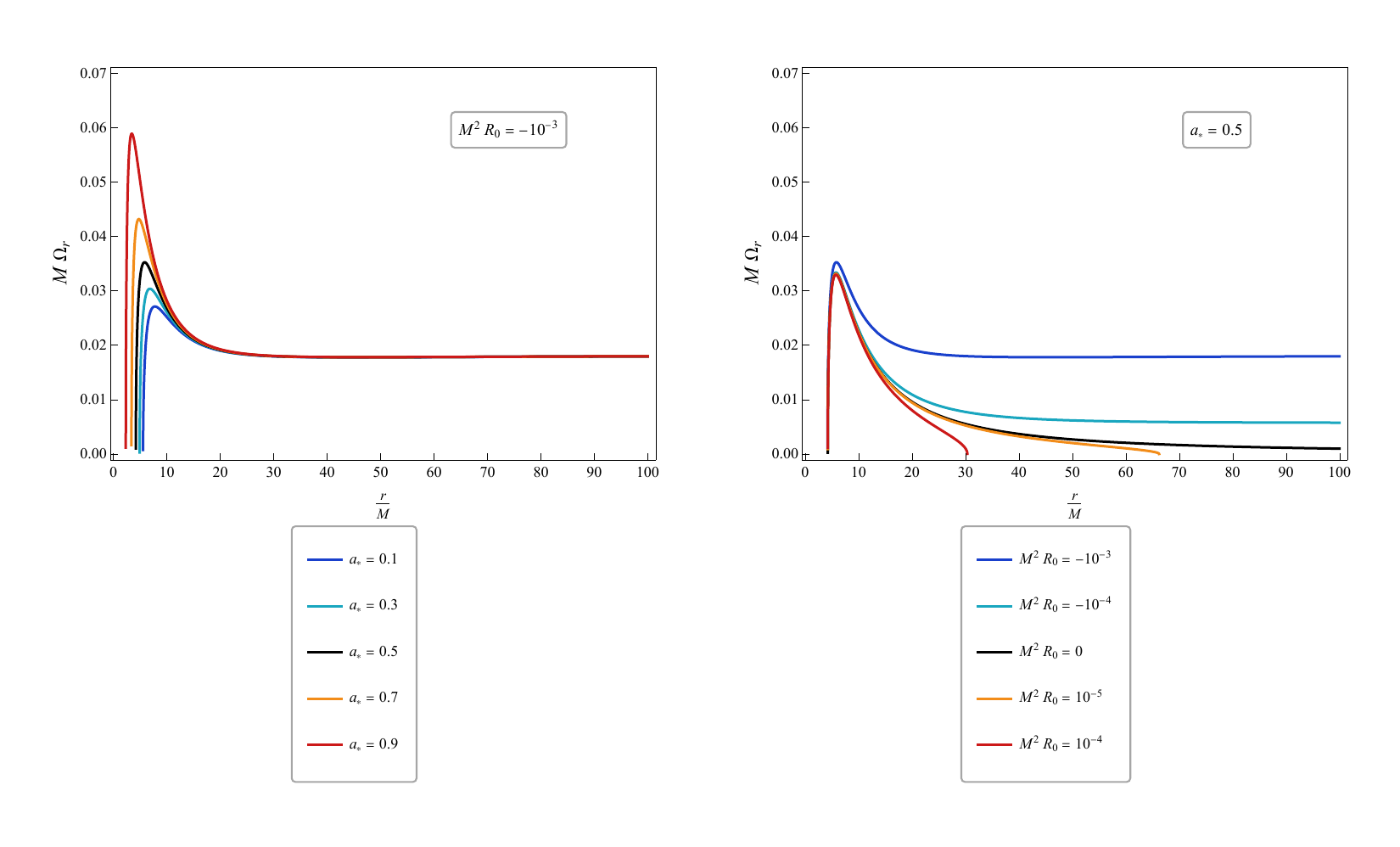}
	\caption{Dimensionless radial epicyclic frequency $M\Omega_r$ in the
rotating background as a function of the dimensionless radial coordinate
$r/M$, computed from Eq.~\eqref{app-kerr-omega-r}. Each curve is shown
only within its domain of radial stability. In the left panel, the
dimensionless curvature is fixed at $R_0M^2=-10^{-3}$, while representative
values of the dimensionless spin $a_*=a/M$ are considered. Increasing the
prograde spin moves the ISCO inward and raises the maximum of
$M\Omega_r$. At large radii, the curves approach
$M\Omega_r\rightarrow\sqrt{|R_0M^2|/3}/\Xi$, where
$\Xi=1+a_*^2R_0M^2/12$. For the parameters shown, $\Xi$ is close
to unity, so the residual spin dependence of the asymptotic plateau
is too small to be readily distinguished in the figure.
In the right panel, the spin is fixed at $a_*=0.5$, and representative
values of $R_0M^2$ are considered. The positive-curvature curves
vanish at both the ISCO and the OSCO. The $R_0M^2=0$ Kerr limit
is included in the right panel for comparison..}
\label{fig:kerr-radial}
\end{figure*}

\begin{figure*}[htbp]
\centering
\includegraphics[width=1\textwidth]{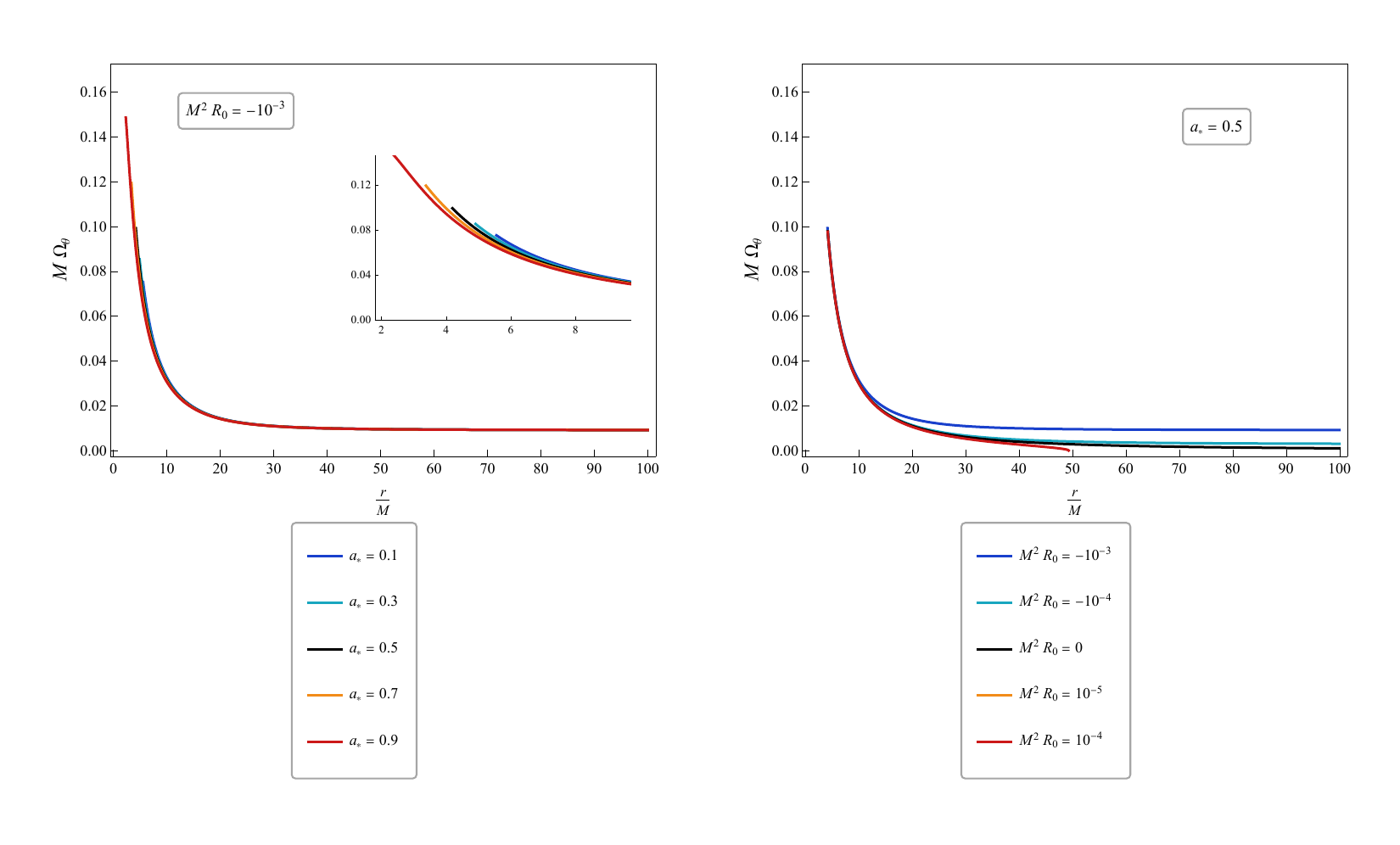}
	\caption{Dimensionless vertical epicyclic frequency $M\Omega_\theta$
in the rotating background as a function of the dimensionless radial
coordinate $r/M$, computed from
Eq.~\eqref{app-kerr-omega-theta}, with the same values of $a_*=a/M$ and
$R_0M^2$ as in Fig.~\ref{fig:kerr-radial}. Each curve begins at its
corresponding ISCO. In the left panel, at fixed $r/M$, increasing the
prograde spin lowers $M\Omega_\theta$, owing to the spin-dependent term
in Eq.~\eqref{app-kerr-omega-theta}, opposite to its effect on
$M\Omega_r$, while extending the stable-orbit domain inward. In the
right panel, the dimensionless curvature $R_0M^2$ controls the
large-radius behavior, and the $R_0M^2>0$ curves terminate at their
corresponding dimensionless static radii $r_s/M$.}
\label{fig:kerr-vertical}
\end{figure*}

We first collect the closed-form expressions for equatorial circular geodesics of the line element \eqref{eq:kerr-ads-metric}. Defining
\begin{equation}\label{eq:J}
\mathcal{J}(r)=\sqrt{\frac{M}{r^{3}}-\frac{R_0}{12}},
\end{equation}
the
coordinate-time epicyclic frequencies read
\begin{equation}\label{app-kerr-omega-theta}
\Xi^{2}\left(1+a\mathcal{J}\right)^{2}\Omega_\theta^{2}
=\mathcal{J}^{2}\left(1+a\mathcal{J}\right)^{2}
-\frac{Ma}{r^{3}}
\left(a\mathcal{J}^{2}+6\mathcal{J}-\frac{3a}{r^{2}}\right),
\end{equation}
\begin{equation}\label{app-kerr-omega-r}
\Xi^{2}\left(1+a\mathcal{J}\right)^{2}\Omega_r^{2}
=4\mathcal{J}^{2}\left(1+a\mathcal{J}\right)^{2}
-\frac{M}{r^{3}}
\left[3+15\mathcal{J}^{2}r^{2}+7a^{2}\mathcal{J}^{2}
+\frac{3a^{2}}{r^{2}}-\frac{9M}{r}-\frac{3Ma^{2}}{r^{3}}\right].
\end{equation}
Both right-hand sides are polynomial in $\mathcal{J}$, and (i)~for $a\to0$
(where $\Xi\to1$),
Eqs.~\eqref{app-kerr-omega-theta}--\eqref{app-kerr-omega-r} reduce
exactly to the static results
\eqref{static-Omega-phi-theta}--\eqref{static-Omega-r}, (ii)~for
$R_0\to0$ and $\Xi\to1$,
they reduce to the standard Kerr epicyclic frequencies
\begin{equation}\label{kerr-limits}
\Omega_r^{2}=\Omega_\phi^{2}
\left(1-\frac{6M}{r}+\frac{8a\sqrt{M}}{r^{3/2}}
-\frac{3a^{2}}{r^{2}}\right),
\qquad
\Omega_\theta^{2}=\Omega_\phi^{2}
\left(1-\frac{4a\sqrt{M}}{r^{3/2}}+\frac{3a^{2}}{r^{2}}\right).
\end{equation}
Evaluating Eq.~\eqref{orbital-frequency-general} on the equatorial
plane of the metric \eqref{eq:kerr-ads-metric} gives the compact result
\cite{Stuchlik:2003dt}
\begin{equation}\label{kerr-ds-orbital-frequency}
\Omega_\phi^{(\pm)}=\pm\frac{\mathcal{J}}{1\pm a\,\mathcal{J}},
\end{equation}
where the upper (lower) sign refers to the corotating (counterrotating)
family. In the static limit, $\Omega_\phi^2\to\mathcal{J}^2$, recovering
Eq.~\eqref{static-Omega-phi-theta}, while for $R_0=0$ one obtains the
familiar Kerr expression
\[
\Omega_\phi^{(\pm)}
=
\frac{\pm\sqrt{M}}{r^{3/2}\pm a\sqrt{M}}.
\] As in the static
case, circular geodesics exist only where $\mathcal{J}^2\geq0$, i.e.\
inside the static radius \eqref{static-radius} when $R_0>0$.
The specific energy and angular momentum follow from
Eq.~\eqref{EL-circular}, and the radial and vertical epicyclic
frequencies from Eq.~\eqref{frequency-original-potential}. Introducing the timelike-existence function
\begin{equation}\label{kerr-C-def}
\mathcal{C}(r):=1-\frac{3M}{r}+2a\mathcal{J}-\frac{a^{2}R_0}{12},
\end{equation}
which must be positive for a circular orbit to be timelike, one finds
the remarkably compact closed forms
\begin{equation}\label{kerr-EL}
E=\frac{1-\dfrac{2M}{r}-\dfrac{R_0}{12}\left(r^{2}+a^{2}\right)
+a\mathcal{J}}{\Xi\,\sqrt{\mathcal{C}}},
\qquad
L=\frac{\left(r^{2}+a^{2}\right)\mathcal{J}
-a\left[\dfrac{2M}{r}+\dfrac{R_0}{12}\left(r^{2}+a^{2}\right)\right]}
{\Xi\,\sqrt{\mathcal{C}}},
\qquad
u^{t}=\frac{\Xi\,(1+a\mathcal{J})}{\sqrt{\mathcal{C}}},
\end{equation}
which generalize the familiar Kerr results (recovered for $R_0\to0$),
and reduce to Eq.~\eqref{static-EL} for $a\to0$. The radial and
vertical epicyclic frequencies then follow from Eqs.~\eqref{app-kerr-omega-theta}--\eqref{app-kerr-omega-r}. Their behavior is displayed in
Figs.~\ref{fig:kerr-radial}--\ref{fig:kerr-vertical}.

\section{Epicyclic-frequency ratios and resonance branches}
\label{sec:resonances}

Twin high-frequency QPOs, when present, are observed in stellar-mass BH binaries with an upper-to-lower frequency ratio clustering near
$3 : 2$
\cite{Remillard:2006fc,Motta:2014gsa,Abramowicz:2001bi}. In the
epicyclic-resonance interpretation, this clustering is attributed to a
nonlinear nonautonomous coupling between the radial and vertical
epicyclic modes, which becomes efficient only near radii where their
frequencies satisfy a low-order rational relation: internal parametric resonance at $\Omega_\theta/\Omega_r=3 : 2$ is
generically the strongest, since it corresponds to the lowest-order
term allowed by the (approximately) quadratic form of the coupling,
while forced resonances driven by orbital eccentricity or
disk-tilt harmonics can additionally excite the $2 : 1$ and $3 : 1$
ratios
\cite{Abramowicz:2001bi,Kluzniak:2001ar,Rebusco:2004ba,Torok:2005ut}.
Under this interpretation, the resonance radius is a genuine
geometric probe: for a source of known mass and spin, the observed frequency pair fixes $r$ 
and constant background curvature shifts that radius in a
calculable way. Because $\Omega_\theta/\Omega_r$ is invariant under a common
constant rescaling of the stationary time coordinate, the
resonance radii determined below do not depend on that
normalization. This invariance does not, by itself, specify how
the associated oscillations produce the frequencies measured
in a distant light curve.
We will start our analysis considering the static scenario.

\subsection{Static background}
\label{sec:resonances-static}

\begin{figure*}[htbp]
\centering
\includegraphics[width=1\textwidth]{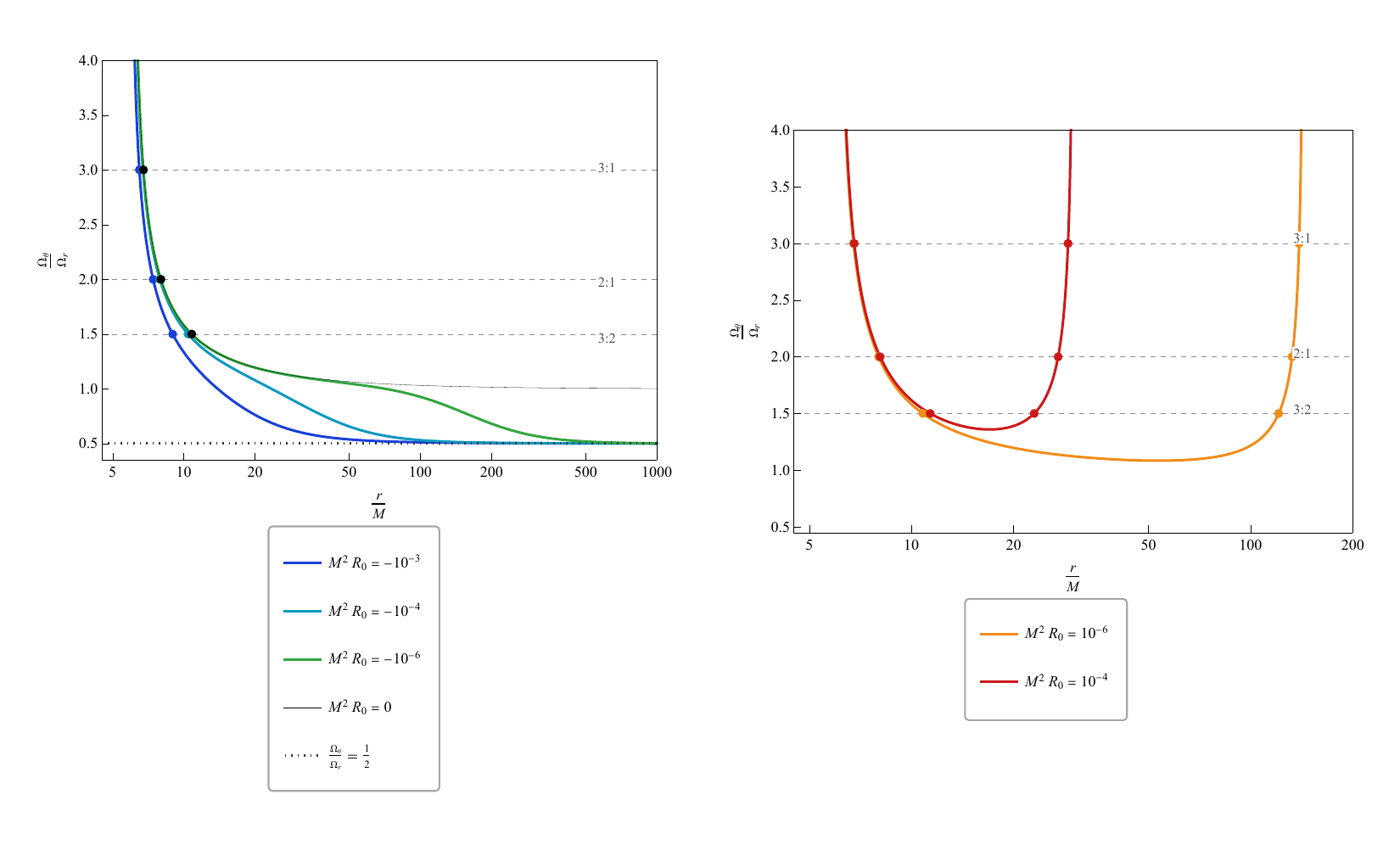}
	\caption{Frequency ratio $\Omega_\theta/\Omega_r$ in the static
background as a function of the dimensionless radial coordinate $r/M$
on a logarithmic scale. In the left panel, for $R_0M^2\leq0$, each
curve diverges at its ISCO and decreases monotonically. The
asymptotically flat case, $R_0M^2=0$ (thin black line), approaches
unity, whereas every negative-curvature curve crosses unity at the
dimensionless hierarchy-inversion radius $r_\times/M$ and approaches
the universal asymptotic value $1/2$ (dotted line), independently of
the magnitude of $R_0M^2$. In the right panel, for $R_0M^2>0$, the
curves diverge at both the ISCO and the OSCO and possess a minimum
between them, allowing each resonance condition to be realized at two
distinct radii. In both panels, the horizontal gray lines mark the
ratios $3{:}2$, $2{:}1$, and $3{:}1$, while the dots indicate the
corresponding resonance radii listed in
Table~\ref{tab:static-resonances}.}
\label{fig:static-ratio}
\end{figure*}

From Eqs.~\eqref{static-Omega-phi-theta}--\eqref{static-Omega-r}, the
ratio $\mathcal{R}(r)$ in the static background is given by
\begin{equation}\label{static-ratio}
\mathcal{R}(r)=\frac{\Omega_\theta}{\Omega_r}
=\sqrt{\frac{\dfrac{M}{r^3}-\dfrac{R_0}{12}}
{\dfrac{M}{r^3}-\dfrac{6M^2}{r^4}+\dfrac{5MR_0}{4r}-\dfrac{R_0}{3}}}\,,
\end{equation}
where the square root is evaluated only in the domain in which both squared frequencies are positive and its qualitative behavior is controlled by the sign of $R_0$. For the asymptotically flat case ($R_0=0$), we note that
$\mathcal{R}=(1-6M/r)^{-1/2}$ decreases monotonically from infinity at
the ISCO to unity at large radii; each resonance condition
$\Omega_\theta/\Omega_r=n : m>1$ is met at exactly one radius. For a negative curvature ($R_0<0$), $\mathcal{R}$ again diverges at the ISCO and decreases monotonically,
but its asymptotic value is now
\begin{equation}\label{ratio-asymptotics}
\lim_{r\to+\infty}\mathcal{R}(r)=\lim_{r\to+\infty}\left(\frac{\Omega_\theta}{\Omega_r}\right)
=\frac{1}{2}.
\end{equation}
Independently of the magnitude of $R_0$, at large radii the radial
oscillation is faster than the vertical one, a qualitative inversion of the near-BH hierarchy with no analogue in asymptotically
flat spacetimes. Each of the ratios $3 : 2$, $2 : 1$, and $3 : 1$ is
therefore realized at exactly one radius, and in addition there exists
a characteristic radius $r_\times$ where the hierarchy inverts,
$\Omega_\theta(r_\times)=\Omega_r(r_\times)$, given by the real
positive root of
\begin{equation}\label{crossing-radius}
\frac{R_0}{4}\,r^4-\frac{5MR_0}{4}\,r^3+6M^2=0
\qquad\Longrightarrow\qquad
r_\times\simeq\left(\frac{24M^2}{|R_0|}\right)^{1/4}
\quad(|R_0|M^2\ll1),
\end{equation}
and beyond $r_\times$, we have
$\Omega_r>\Omega_\phi=\Omega_\theta$. Finally, for a positive curvature ($0<R_0<R_0^{\rm crit}$), the radial epicyclic frequency vanishes at both boundaries of the stable annulus. Consequently, $\mathcal{R} \to +\infty$ as $r \to r^{+}_{\tiny{\mbox{ISCO}}}$ and as $r \to r^{-}_{\tiny{\mbox{OSCO}}}$. Between these boundaries, $\mathcal{R}$ attains a finite minimum $\mathcal{R}_{\tiny{\mbox{min}}}(R_0)$. Therefore, a resonance condition
\[
\mathcal{R}=\frac{n}{m},
\]
possesses two distinct stable radial solutions whenever
\[
\frac{n}{m}>\mathcal{R}_{\tiny{\mbox{min}}}(R_0).
\]
These solutions disappear for larger curvature. This double-valued resonance structure is a characteristic geometric feature of the positive constant-curvature branch, although it is not a theory-specific signature of metric $f(R)$ gravity.
Figure~\ref{fig:static-ratio} shows $\Omega_\theta/\Omega_r$ for both
signs of $R_0$ together with the location of the $3{:}2$, $2{:}1$, and
$3{:}1$ branches, and Table~\ref{tab:static-resonances} lists the
corresponding radii.

\begin{table}[htbp]
\centering
\caption{Dimensionless resonance radii in the static background,
obtained from Eq.~\eqref{static-ratio}. For $R_0M^2>0$, each resonance
condition is realized on an inner (ISCO-side) branch and an outer
(OSCO-side) branch. The asymptotically flat values
$r_{3:1}/M=6.75$, $r_{2:1}/M=8$, and $r_{3:2}/M=10.8$ are included
for reference.}
\label{tab:static-resonances}
\begin{tabular}{lcccc}
\hline\hline
$R_0M^2$ & Branch & $r_{3:1}/M$ & $r_{2:1}/M$ & $r_{3:2}/M$ \\
\hline
$-10^{-3}$ & --    & 6.488  & 7.418  & 8.971  \\
$-10^{-6}$ & --    & 6.750  & 7.999  & 10.796 \\
$0$        & --    & 6.750  & 8.000  & 10.800 \\
$+10^{-6}$ & inner & 6.750  & 8.001  & 10.804 \\
           & outer & 138.97 & 132.48 & 120.95 \\
$+10^{-4}$ & inner & 6.783  & 8.090  & 11.358 \\
           & outer & 28.95  & 27.07  & 23.01  \\
\hline\hline
\end{tabular}
\end{table}

\subsection{Rotating background}
\label{sec:resonances-kerr}
\begin{figure*}[htbp]
\centering
\includegraphics[width=1\textwidth]{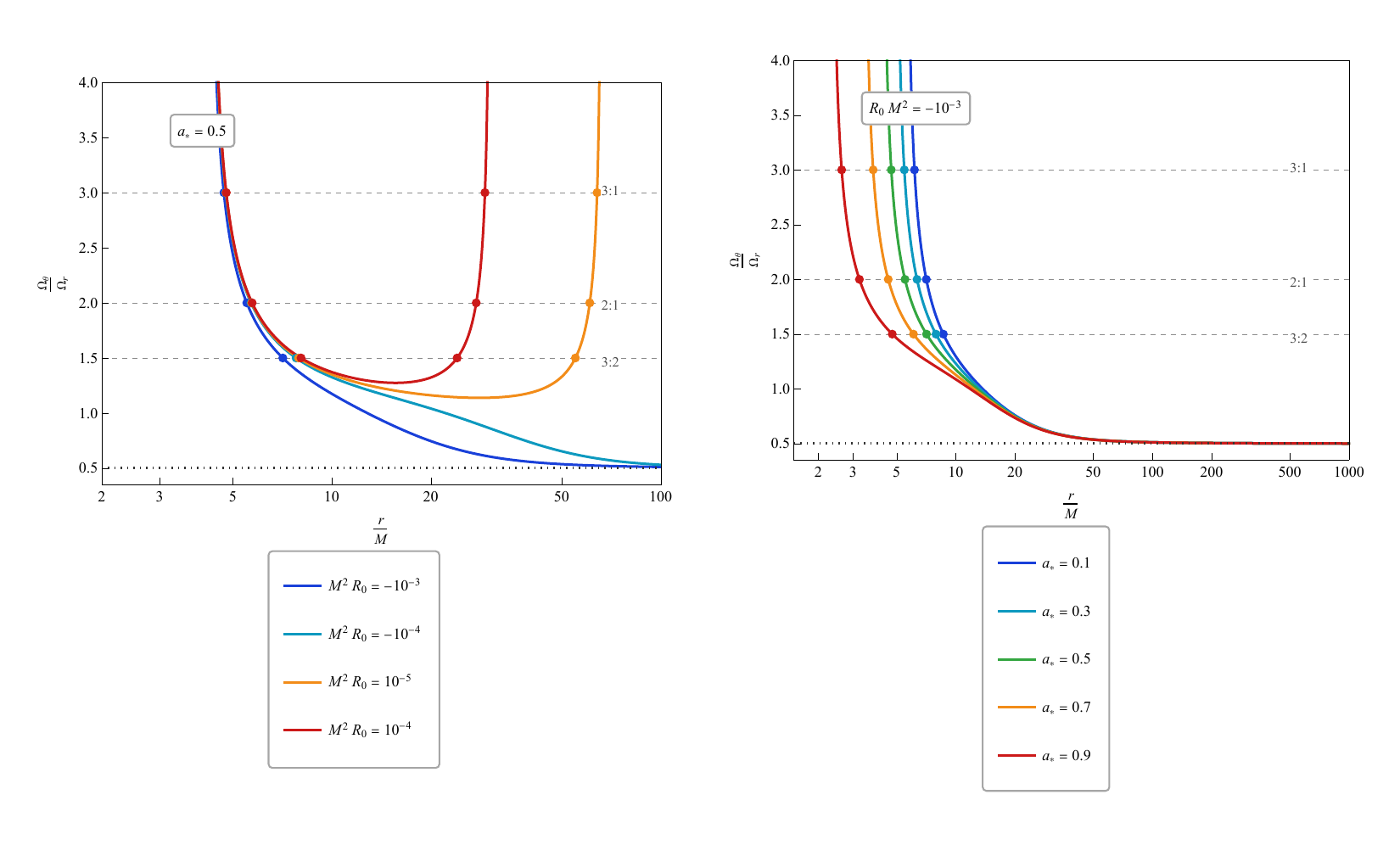}
	\caption{Frequency ratio $\Omega_\theta/\Omega_r$ in the rotating
background for the corotating family, shown as a function of the
dimensionless radial coordinate $r/M$ on a logarithmic scale. The dots
mark the $3{:}2$, $2{:}1$, and $3{:}1$ resonance radii listed in
Tables~\ref{tab:kerr-resonances-a}--\ref{tab:kerr-resonances-R0}.
In the left panel, the dimensionless spin is fixed at $a_*=a/M=0.5$.
For the positive-curvature values shown, the frequency ratio
diverges at the ISCO and OSCO and has a minimum between them,
as in the static case [see the right panel of
Fig.~\ref{fig:static-ratio}]. Both resonance branches are present
for the selected ratios because these curvature values lie below
the corresponding resonance-merger thresholds given in Eq.~(63).
In the right panel, the dimensionless curvature is fixed at
$R_0M^2=-10^{-3}$. Increasing $a_*$ shifts every resonance branch
toward smaller values of $r/M$ without changing the large-radius
asymptote $1/2$
[see Eq.~\eqref{ratio-asymptotics}].}
\label{fig:kerr-ratio}
\end{figure*}

In the rotating case, the ratio $\Omega_\theta/\Omega_r$, given in Eqs.~\eqref{app-kerr-omega-theta}--\eqref{app-kerr-omega-r}, inherits the qualitative structure of the static background. For $R_0\leq0$, the ratio decreases from its divergence at the ISCO, whereas for $0<R_0<R_0^{\rm crit}$, it develops a divergence--minimum--divergence profile between the ISCO and the OSCO. Increasing the corotating spin moves the ISCO and the associated resonance radii inward, as in Kerr. For $R_0M^2=-10^{-3}$, the $3:2$ resonance radius decreases from
$r_{3:2}/M=8.645$ at $a_\ast=0.1$ to
$r_{3:2}/M=4.753$ at $a_\ast=0.9$. (see Table~\ref{tab:kerr-resonances-a}). The $3 : 1$ and $2 : 1$ radii follow the same pattern. For positive curvature, $R_0>0$, prograde rotation also changes the maximum curvature for which each pair of resonance branches exists. For $a_\ast=0.5$, the inner and outer resonance branches disappear
at approximately
\begin{equation}
 \left(R_0M^2\right)^{(3:2)}_{\rm merge}
 \simeq 5.3\times10^{-4},
 \qquad
 \left(R_0M^2\right)^{(2:1)}_{\rm merge}
 \simeq 2.1\times10^{-3},
 \qquad
 \left(R_0M^2\right)^{(3:1)}_{\rm merge}
 \simeq 4.8\times10^{-3}.
\end{equation}
 Thus, prograde rotation enlarges the curvature interval over which the OSCO-side resonance branch survives. Finally, since
$\Omega_\phi\neq\Omega_\theta$ once $a\neq0$, the single crossing
radius $r_\times$, obtained in Sec.~\ref{sec:resonances-static},   where
$\Omega_r=\Omega_\theta$ and the near-horizon hierarchy
$\Omega_\phi>\Omega_\theta>\Omega_r$ inverts to
$\Omega_r>\Omega_\phi>\Omega_\theta$, becomes spin-dependent. For $R_0M^2=-10^{-3}$, the hierarchy-inversion radius changes from
$r_\times/M\simeq13.91$ at $a_\ast=0$ to
$r_\times/M\simeq11.83$ at $a_\ast=0.9$. The inversion is thus
predominantly a property of the background curvature, with rotation
acting as a secondary correction. Figure~\ref{fig:kerr-ratio} displays the ratio for representative spins
and curvatures, and Tables~\ref{tab:kerr-resonances-a}--\ref{tab:kerr-resonances-R0} collect the resonance radii.

\begin{table}[htbp]
\centering
\caption{Effect of the dimensionless spin $a_*=a/M$ on the
dimensionless resonance radii for $R_0M^2=-10^{-3}$ in the corotating
family. The inward shift with increasing spin, already present in the
Kerr limit $R_0M^2=0$, is monotonic and comparable in magnitude to the
shift produced by the curvature values considered in
Table~\ref{tab:kerr-resonances-R0}.}
\label{tab:kerr-resonances-a}
\begin{tabular}{lccc}
\hline\hline
$a_*$ & $r_{3:1}/M$ & $r_{2:1}/M$ & $r_{3:2}/M$ \\
\hline
0.1 & 6.166 & 7.082 & 8.645 \\
0.3 & 5.473 & 6.351 & 7.930 \\
0.5 & 4.699 & 5.518 & 7.100 \\
0.7 & 3.801 & 4.533 & 6.094 \\
0.9 & 2.629 & 3.236 & 4.753 \\
\hline\hline
\end{tabular}
\end{table}
 
\begin{table}[htbp]
\centering
\caption{Dimensionless resonance radii in the rotating background for
$a_*=a/M=0.5$ and representative values of the dimensionless curvature
$R_0M^2$ in the corotating family. For $R_0M^2>0$, the inner
(ISCO-side) and outer (OSCO-side) branches are listed separately. The
outer branch survives only below the spin-dependent merger curvature
identified in this work; for the $3{:}2$ resonance pair at $a_*=0.5$,
this condition is approximately $R_0M^2\lesssim5.3\times10^{-4}$.}
\label{tab:kerr-resonances-R0}
\begin{tabular}{lcccc}
\hline\hline
$R_0M^2$ & Branch & $r_{3:1}/M$ & $r_{2:1}/M$ & $r_{3:2}/M$ \\
\hline
$-10^{-3}$ & --    & 4.699 & 5.518 & 7.100 \\
$-10^{-4}$ & --    & 4.764 & 5.685 & 7.792 \\
$+10^{-5}$ & inner & 4.773 & 5.710 & 7.933 \\
           & outer & 64.07 & 60.87 & 55.04 \\
$+10^{-4}$ & inner & 4.780 & 5.731 & 8.068 \\
           & outer & 29.22 & 27.49 & 24.05 \\
\hline\hline
\end{tabular}
\end{table}

\section{Nodal precession, global rigid precession, and viscous
alignment}
\label{sec:precession}

\subsection{Nodal precession}
\label{sec:nodal}

\begin{figure*}[htbp]
\centering
\includegraphics[width=1\textwidth]{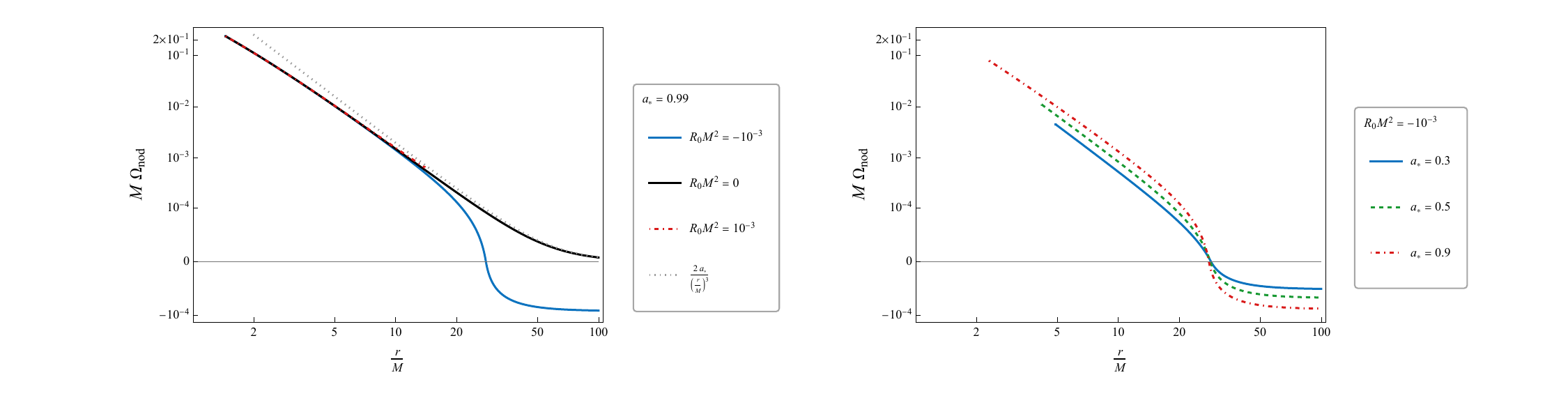}
	\caption{Dimensionless signed nodal-precession frequency
$M\Omega_{\rm nod}$ for the corotating family as a function of the
dimensionless radial coordinate $r/M$. A symmetric-logarithmic vertical
scale is used so that both positive and negative values are visible.
In the left panel, the dimensionless spin is fixed at $a_*=a/M=0.99$,
and the curvature values are
$R_0M^2=\{-10^{-3},\,0,\,10^{-3}\}$. The dotted gray curve represents
the dimensionless form of the Lense--Thirring approximation given in
Eq.~\eqref{lense-thirring-limit}. The positive-curvature profile
terminates at the OSCO and lies above the Kerr profile at large radii,
whereas the negative-curvature profile crosses zero at
$r/M\simeq r_{\rm rev}/M$ and becomes retrograde beyond this radius.
In the right panel, the dimensionless curvature is fixed at
$R_0M^2=-10^{-3}$, and the dimensionless spins are
$a_*=\{0.3,\,0.5,\,0.9\}$. The reversal radius $r_{\rm rev}/M$ is
nearly independent of $a_*$, as predicted by the slow-rotation
approximation.}
\label{fig:nodal}
\end{figure*}

 In the rotating background, the loss of degeneracy between azimuthal
and vertical motion produces a precession of the orbital plane of a
slightly tilted circular orbit at the signed nodal frequency
\begin{equation}\label{nodal-definition}
\Onod(r)=\Omega_\phi(r)-\Omega_\theta(r),
\end{equation}
which vanishes identically in the static limit
\eqref{static-Omega-phi-theta} and is therefore a genuine
frame-dragging effect. In the weak-field, slow-rotation, and
small-curvature limit, $\Onod$ reduces to the standard Lense--Thirring
expression:
\begin{equation}\label{lense-thirring-limit}
\Onod\longrightarrow\frac{2Ma}{r^{3}},
\mbox{ for }
\frac{M}{r}\ll1,\quad \frac{a}{M}\ll1,\quad |R_0|\,r^2\ll1.
\end{equation}
Expanding the exact expressions of
Eqs.~\eqref{kerr-ds-orbital-frequency} and \eqref{app-kerr-omega-theta} to first order in the spin, we
obtain the closed-form generalization
\begin{equation}\label{nodal-linear}
\Onod=a\left(\frac{2M}{r^{3}}+\frac{R_0}{12}\right)
+\mathcal{O}(a^{2}),
\end{equation}
which makes the curvature correction to the Lense--Thirring precession
explicit; the constant background curvature contributes the
$r$-independent term $aR_0/12$, so that de~Sitter curvature enhances the nodal precession at large radii, while
anti-de~Sitter curvature opposes it and, remarkably, reverses its
sign beyond the radius
\begin{equation}\label{nodal-reversal}
r_{\rm rev}=\left(\frac{24M}{|R_0|}\right)^{1/3}
\qquad(R_0<0),
\end{equation}
where the orbital plane of a tilted orbit precesses opposite to the BH spin. The slow-rotation estimate \eqref{nodal-reversal} is
remarkably accurate even at high spin; for $R_0M^2=-10^{-3}$, the nodal-precession reversal occurs near
$r_{\rm rev}/M\simeq28.8$. Its weak dependence on spin is illustrated
by $r_{\rm rev}/M\simeq28.6$ for $a_\ast=0.3$ and
$r_{\rm rev}/M\simeq28.0$ for $a_\ast=0.9$. Figure~\ref{fig:nodal} shows $\Onod$ for
representative spins and curvatures. Three features are worth
emphasizing. First, at a fixed radius in the inner region, the
magnitude of $\Onod$ grows with the spin, as in the Kerr situation. Second, at
fixed spin, the curvature modifies the profile predominantly at
large radii, where the background terms $\propto R_0$ in
$\Omega_\phi$ and $\Omega_\theta$ become comparable to the Keplerian
ones; for $R_0>0$ the profile terminates at the OSCO. Third, the
sign reversal \eqref{nodal-reversal} in the anti-de~Sitter branch
implies that a rigidly precessing flow extending across $r_{\rm rev}$
receives competing torques from its inner and outer regions.
 
\subsection{Global rigid precession of a tilted flow}
\label{sec:rigid}

\begin{figure*}[htbp]
\centering
\includegraphics[width=0.6\textwidth]{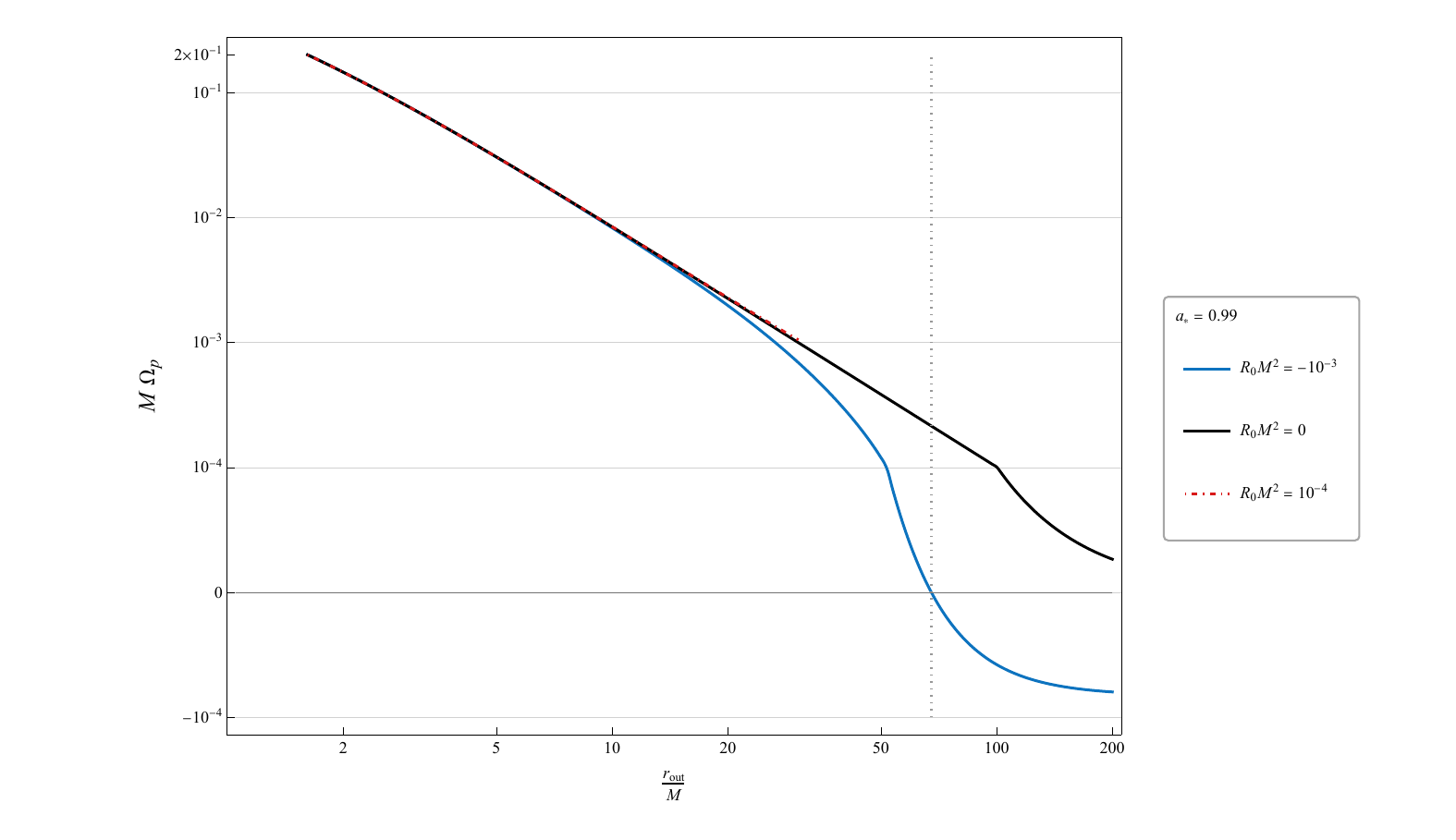}
\caption{Dimensionless global rigid-precession frequency
$M\Omega_p$, defined in Eq.~\eqref{rigid-precession}, as a function of
the dimensionless outer truncation radius $r_{\rm out}/M$. A
symmetric-logarithmic vertical scale is used to display both signs.
The dimensionless spin is fixed at $a_*=a/M=0.99$, and the curvature
values are $R_0M^2=\{-10^{-3},\,0,\,10^{-4}\}$. For
$R_0M^2>0$, the curve terminates at the OSCO, in accordance with
Eq.~\eqref{outer-truncation}. The $R_0M^2<0$ curve crosses zero at
$r_{\rm out}/M\simeq67.7$ (dotted line), well beyond the local
nodal-precession reversal radius $r_{\rm rev}/M\simeq28.8$ shown in
Fig.~\ref{fig:nodal}.}
\label{fig:rigid-precession}
\end{figure*}
 
If the angular momentum of a geometrically thick inner flow is
misaligned with the BH spin and the warp-communication time is short
compared with the precession time, the flow precesses approximately as
a rigid body at the angular-momentum-weighted frequency
\cite{Ingram:2009vm,Ingram:2011km}
\begin{equation}\label{rigid-precession}
\Omega_p=
\frac{\displaystyle\int_{r_{\rm in}}^{r_{\rm out}}
\Onod(r)\,\mathcal{L}(r)\,2\pi r\,dr}
{\displaystyle\int_{r_{\rm in}}^{r_{\rm out}}
\mathcal{L}(r)\,2\pi r\,dr},
\qquad
\mathcal{L}(r)=\Sigma(r)\,\Omega_\phi(r)\,r^{2},
\end{equation}
where $\Sigma(r)$ is the surface-density profile. Following
Refs.~\cite{Ingram:2009vm,Ingram:2011km} we adopt the convention:
\begin{equation}\label{surface-density}
\Sigma(r)=\Sigma_0\,r^{-p}\left(1-\sqrt{\frac{r_{\rm in}}{r}}\right)^{p},
\end{equation}
with $p=3/5$, appropriate for a radiation-pressure-dominated
Shakura--Sunyaev flow with viscosity proportional to the gas pressure,
and we set the inner edge at the marginally stable orbit,
$r_{\rm in}=\risco$. The choice of the outer edge deserves comment: in
the asymptotically flat and anti-de~Sitter branches, $r_{\rm out}$ is a
free phenomenological parameter set by the truncation of the thick
flow, whereas for $R_0>0$ the OSCO provides a natural upper bound,
\begin{equation}\label{outer-truncation}
r_{\rm out}\leq\rosco ,
\end{equation}
since no stable circular motion exists beyond it. The positive-curvature
branch thus removes, in part, the arbitrariness of the outer truncation
inherent to the rigid-precession prescription. The normalization
$\Sigma_0$ cancels in Eq.~\eqref{rigid-precession}.

We evaluate Eq.~\eqref{rigid-precession} numerically using the frequencies in Eqs.~\eqref{kerr-ds-orbital-frequency} and \eqref{app-kerr-omega-theta} and adaptive Gaussian quadrature. Figure~\ref{fig:rigid-precession} shows the dimensionless global
precession frequency $M\Omega_p$ as a function of $r_{\rm out}/M$
for $a_\ast=0.99$ and representative values of $R_0M^2$ on a symmetric logarithmic vertical scale. For $R_0\geq0$, the global precession frequency remains positive and decreases as $r_{\rm out}$ grows because progressively slower-precessing material is included in the weighted average. For $R_0<0$, the local sign reversal of $\Onod$ eventually propagates to the global frequency. For $R_0M^2=-10^{-3}$, the global precession frequency crosses zero
at $r_{\rm out}/M\simeq67.7$, substantially beyond the local
nodal-precession reversal radius $r_{\rm rev}/M\simeq28.8$. The difference arises because the global frequency contains contributions from the entire precessing region. Thus, a flow extending beyond $r_{\rm rev}$ may still precess globally in the prograde direction until the retrograde torque from the outer region becomes sufficiently large to cancel the inner prograde contribution.

\subsection{Viscous alignment timescale}
\label{sec:alignment}
\begin{figure*}[htbp]
\centering
\includegraphics[width=0.6\textwidth]{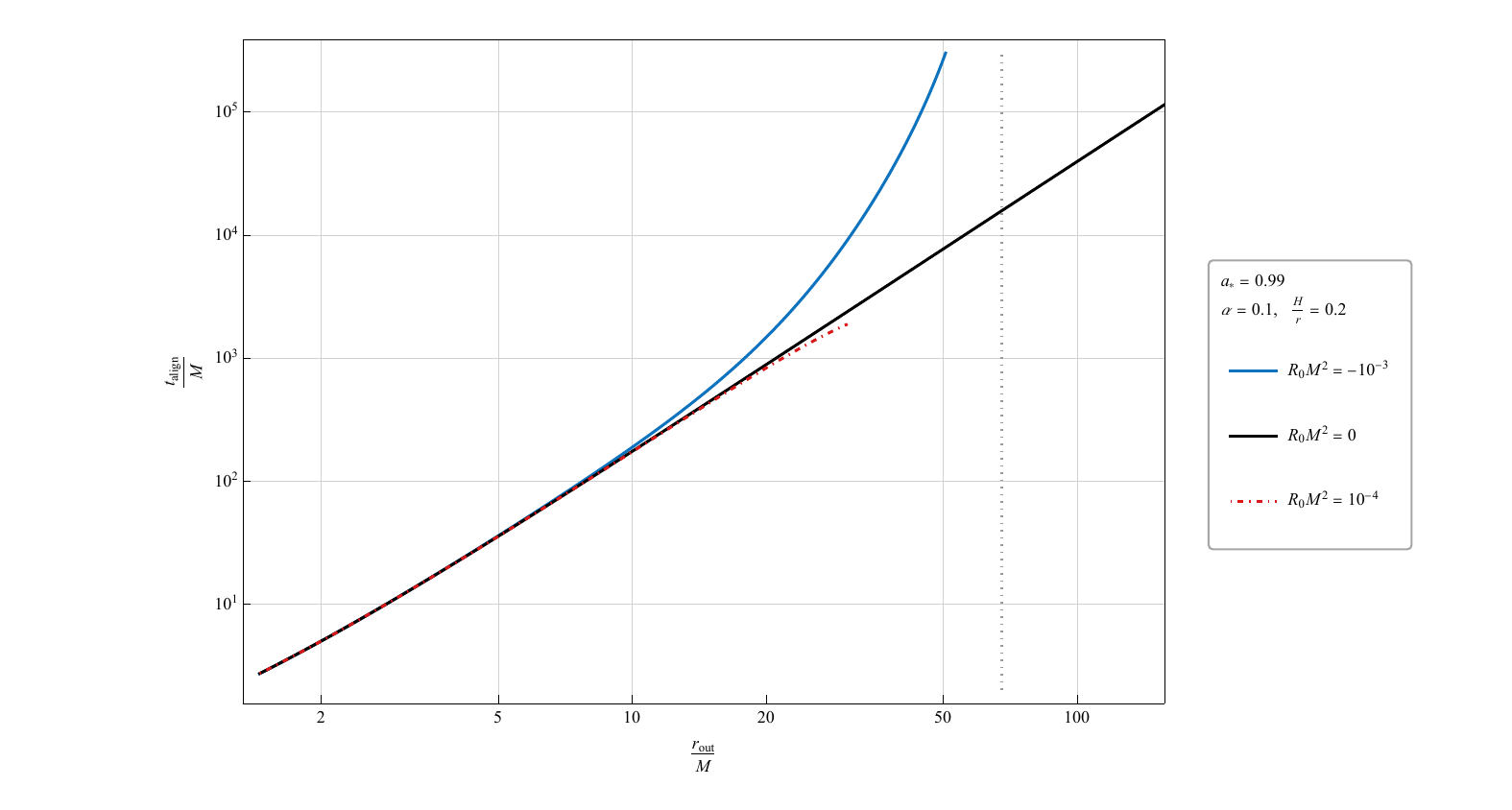}
\caption{Dimensionless viscous alignment timescale
$t_{\rm align}/M$, defined in Eq.~\eqref{alignment-timescale}, as a
function of the dimensionless outer truncation radius $r_{\rm out}/M$.
A logarithmic vertical scale is used. The adopted parameters are
$\alpha=0.1$, $H/r=0.2$, $a_*=a/M=0.99$, and
$R_0M^2=\{-10^{-3},\,0,\,10^{-4}\}$. For the negative-curvature
case, $t_{\rm align}/M$ diverges at
$r_{\rm out}/M\simeq67.7$ (dotted line), where the global precession
frequency $M\Omega_p$ crosses zero
[see Fig.~\ref{fig:rigid-precession}]. The plotted negative-curvature
branch approaches this divergence from smaller outer radii. Away from
the divergence, positive curvature systematically shortens the
alignment timescale relative to negative curvature, consistently with
the corresponding enhancement of $|M\Omega_p|$.}
\label{fig:alignment}
\end{figure*}
Dissipation within the tilted flow damps the precession and drives the
angular momentum of the flow toward alignment with the BH spin. For a
disk of approximately constant aspect ratio $H/r$ and effective
viscosity parameter $\alpha$, the alignment timescale can be estimated
as \cite{Bate:2000cn}
\begin{equation}\label{alignment-timescale}
t_{\rm align}\simeq\frac{1}{\alpha}
\left(\frac{H}{r}\right)^{2}
\frac{\Omega_\phi}{\Omega_p^{2}},
\end{equation}
evaluated at a characteristic radius of the precessing region, for
which we adopt the outer truncation radius $r_{\rm out}$ itself,
since the surface-density weighting \eqref{surface-density} makes
the outer annulus dominate the angular-momentum budget of the flow.
Since $\Omega_p$ is systematically modified by the background
curvature, so is $t_{\rm align}$:
configurations with larger $|\Omega_p|$ align faster. We stress that
Eq.~\eqref{alignment-timescale} is a local, order-of-magnitude
estimate rather than a solution of the full warped-disk diffusion
equation; in particular, it {formally diverges} as
$\Omega_p\to0$, i.e.\ at the global reversal radius identified in
Sec.~\ref{sec:rigid}, signaling the breakdown of the rigid-precession
picture there rather than a genuine infinite alignment time. Close to
that radius, higher-order (bending-wave or diffusive) torques not
captured by Eq.~\eqref{alignment-timescale} would regularize the
timescale. Figure~\ref{fig:alignment} shows the resulting timescale
for representative parameters. Throughout, we adopt the fiducial
values $\alpha=0.1$ and $H/r=0.2$, typical of geometrically thick,
hot inner flows.

\section{Observational constraints from the QPOs of GRO J1655--40}
\label{sec:qpo_constraints}

The frequency analysis developed in the preceding sections can be confronted
directly with X-ray timing observations.  For this purpose, we consider the microquasar GRO J1655--40, for
which Motta et al.~\cite{Motta:2013wga} identified a simultaneous triplet
consisting of two high-frequency QPOs and a type-C QPO in the
\textit{Rossi X-ray Timing Explorer} data..  The simultaneous detection is particularly
useful because, within the relativistic-precession (RP) prescription, the three
signals probe three independent combinations of the orbital and epicyclic
frequencies at one common emission radius.  We use these data to constrain the
dimensionless constant-curvature parameter
\begin{equation}
 \lambda\equiv R_0M^2,
\end{equation}
jointly with the mass, dimensionless spin, and emission radius.

\subsection{RP identification and observational inputs}

Our observational analysis follows the established use of
relativistic-precession frequencies to constrain non-Kerr
geometries with the GRO J1655--40 triplet
\cite{Bambi2015RPM}. Related applications to
Kerr--Newman--de Sitter and standard and extended
Kerr--de Sitter backgrounds have been presented in
Refs.~\cite{Rink2021QPO,Hesamolhokama2024QPO}.
Here we apply this framework to the neutral constant-curvature
branch and quantify the dependence of the inferred curvature
interval on the curvature prior and the independent
dynamical-mass likelihood.

In the RP model, the upper and lower high-frequency QPOs are identified with
the orbital and periastron-precession frequencies, while the type-C QPO is
identified with the nodal-precession frequency:
\begin{equation}
 \nu_U=\nu_\phi,\qquad
 \nu_L=\nu_\phi-\nu_r,\qquad
 \nu_C=\nu_\phi-\nu_\theta .
 \label{eq:rp_identification}
\end{equation}
To avoid ambiguity in restoring physical units, we define
$\bar\Omega_i=M\Omega_i$ and write
\begin{equation}
 \nu_i=\frac{c^3}{2\pi G M}\,\bar\Omega_i .
 \label{eq:frequency_conversion}
\end{equation}
The frequencies $\Omega_\phi$, $\Omega_r$, and $\Omega_\theta$ are evaluated
from Eqs.~\eqref{kerr-ds-orbital-frequency}, (\ref{app-kerr-omega-r}), and (\ref{app-kerr-omega-theta}), respectively, using the stationary coordinate
time $t$ of Eq.~\eqref{eq:kerr-ads-metric}.  Thus, both epicyclic frequencies contain the factor
$\Xi^{-1}$ required by the adopted convention, whereas
$\Omega_\phi=\mathcal{J}/(1+a\mathcal{J})$.  Restoring physical units does not, by itself, specify the relation
between the stationary coordinate time and the clock associated
with the observed signal. In the present inference, we adopt the
phenomenological identification that the observed QPO periods are
represented by the coordinate-time periods in the specified
stationary and azimuthal convention, with the mass-dependent
conversion given above. No additional frequency-transfer factor
between the modeled region and the observer is fitted.

This identification is an assumption of the observational model.
The constant-curvature geometry is treated as an effective
description of the region supporting the QPO-producing orbits;
we do not construct a global continuation from that region to
the observer. A physical determination of the frequency transfer
would require a specified exterior geometry, observer worldline,
and prescription for the generation and propagation of the
modulated radiation. The recovery of the Kerr limit motivates
our reference convention but does not supply that construction
at nonzero curvature.

Accordingly, the posterior reported below is conditional on this
clock and frame identification, in addition to the geodesic RP
assignments and the dynamical-mass likelihood. Alternative
physical prescriptions for connecting the local model to the
observations can change the inferred curvature interval.
A consistent coordinate transformation of a fixed physical
observer model, by contrast, cannot change its observable
predictions.

For GRO J1655--40, we adopt the simultaneous triplet reported in Ref.~\cite{Motta:2013wga},
\begin{align}
 \nu_U^{\rm obs}&=441\pm2~{\rm Hz},&
 \nu_L^{\rm obs}&=298\pm4~{\rm Hz},\nonumber\\
 \nu_C^{\rm obs}&=17.3\pm0.1~{\rm Hz},
 \label{eq:gro_qpo_data}
\end{align}
together with the independent dynamical mass estimate of Ref.~\cite{Beer:2001cg},
\begin{equation}
 M_{\rm dyn}=5.4\pm0.3\,M_\odot.
 \label{eq:gro_mass_data}
\end{equation}
The parameter vector is
\begin{equation}
 \boldsymbol{\vartheta}
=\left(M/M_\odot,a_\ast,x,\lambda\right),
 \qquad a_\ast\equiv a/M,\quad x\equiv r/M.
 \label{eq:qpo_parameter_vector}
\end{equation}

\subsection{Likelihood, priors, and physical domain}

The three frequency measurements and the dynamical mass estimate are treated
as independent Gaussian constraints.  The likelihood is therefore
\begin{align}
 \mathcal{L}(\boldsymbol{\vartheta})
 &\propto\exp\!\left[-\frac{1}{2}
\chi^2(\boldsymbol{\vartheta})\right],
 \label{eq:qpo_likelihood}\\
 \chi^2(\boldsymbol{\vartheta})
 &=\sum_{j\in\{U,L,C\}}
 \left[\frac{\nu_j^{\rm th}(\boldsymbol{\vartheta})-
 \nu_j^{\rm obs}}{\sigma_j}\right]^2
 +\left[\frac{M/M_\odot-5.4}{0.3}\right]^2.
 \label{eq:qpo_chi2}
\end{align}

The curvature constraint derived below is obtained from the joint
QPO-plus-dynamical-mass posterior. This distinction is important because
the three measured QPO frequencies constrain four model parameters
\begin{equation}
\boldsymbol{\vartheta}
=
\left(
M/M_{\odot},a_{\ast},x,\lambda
\right).
\end{equation}
Consequently, the frequency-only inference is underdetermined: the QPO
triplet constrains three combinations of these parameters but does not
independently determine all four quantities. The Gaussian dynamical-mass
likelihood in the second term of {Eq.~\eqref{eq:qpo_chi2}} supplies the
additional astrophysical information required to obtain the marginalized
constraint on $\lambda=R_{0}M^{2}$.
The primary calculation uses the separable uniform prior
\begin{align}
 4&<M/M_\odot<7,&0&<a_\ast<0.998,\nonumber\\
 1&<x<30,&-0.03&<\lambda<0.03 .
 \label{eq:qpo_priors}
\end{align}
We subsequently repeat the complete inference with
$-0.01<\lambda<0.01$, leaving every other prior, likelihood term, numerical
setting, and physical cut unchanged.  This second calculation is used solely
as a controlled prior-sensitivity test.

The prior support is further restricted to physical, timelike, radially and
vertically stable circular orbits.  In the dimensionless variables of the
sampling calculation, accepted points satisfy
\begin{align}
 \mathcal{J}^2&=x^{-3}-\lambda/12>0,&
 \Xi&=1+a_\ast^2\lambda/12>0,\nonumber\\
 \mathcal{C}&>0,&\Delta_r&>0,\nonumber\\
 \Omega_r^2&>0,&\Omega_\theta^2&>0.
 \label{eq:qpo_physical_cuts}
\end{align}
Since the condition $\Delta_r>0$ alone does not uniquely identify the
domain of outer communication when several positive radial intervals
exist, we additionally calculated the horizon radii for every retained
posterior sample. In terms of the dimensionless variables used in the
inference, the horizon equation can be written as
\begin{equation}
 \bar{\Delta}_r(x)
 \equiv \frac{\Delta_r}{M^2}
 =
 \left(x^2+a_\ast^2\right)
 \left(1-\frac{\lambda x^2}{12}\right)-2x=0.
 \label{eq:qpo_horizon_equation}
\end{equation}
For $\lambda\leq0$, we verified that the sampled QPO orbit satisfies
$x_{\rm QPO}>x_+$, where $x_+$ denotes the event-horizon radius.
For the de~Sitter branch, $\lambda>0$, we verified the stronger
condition
\begin{equation}
 x_+<x_{\rm QPO}<x_c,
 \label{eq:qpo_horizon_domain}
\end{equation}
where $x_c$ denotes the cosmological-horizon radius. All
$4.0\times10^4$ retained samples in each of the baseline,
narrow-curvature-prior, broadened-mass, and QPO-only calculations
satisfy the corresponding horizon-domain condition. In the primary
run, the minimum separations are
\begin{equation}
 \min(x_{\rm QPO}-x_+)=3.132,
 \qquad
 \min_{\lambda>0}(x_c-x_{\rm QPO})=56.741.
\label{eq:qpo_minimum_horizon_separations}
\end{equation}
Across all four calculations, the corresponding conservative minima
are $2.921$ and $51.602$, respectively. The inferred posterior
therefore occupies the exterior stable branch continuously connected
to the Kerr domain and receives no contribution from regions inside
the event horizon or beyond the cosmological horizon.

\subsection{Sampling strategy and convergence}

Posterior sampling was carried out in \textsc{Mathematica} with four
independent Metropolis--Hastings chains.  A pilot stage of
$3.0\times10^4$ steps per chain was used to estimate a multivariate proposal
covariance, which was then held fixed during the production stage.  Each
production chain contained $1.2\times10^5$ states.  The first
$2.0\times10^4$ states of each chain were discarded, and every tenth remaining
state was retained, giving $4.0\times10^4$ posterior samples per run.  The
chains were initialized from dispersed points and used independent random
streams.

The convergence statistics for both curvature priors are collected in
Table~\ref{tab:qpo_diagnostics}.  For the primary run, the four acceptance
fractions range from $0.1564$ to $0.1583$, all split-$\widehat R$ values are
below $1.0004$, and the approximate effective sample sizes lie between 8838
and 10940.  For the narrower-prior run, the acceptance fractions increase to
$0.1729$--$0.1787$, the largest split-$\widehat R$ is $1.0017$, and the
effective sample sizes remain above 7391. Thus, both runs satisfy the commonly adopted convergence requirement
$\widehat{R}<1.01$. The modest
acceptance rate of the primary random-walk chains does not compromise the
inference because the between-chain agreement and effective sample sizes are
both satisfactory.

\begin{table*}[t]
 \centering
 \caption{MCMC diagnostics for the primary inference and the controlled
 prior-sensitivity run.  Entries in the last two columns follow the parameter
 order $(M/M_\odot,a_\ast,x,\lambda)$.}
 \label{tab:qpo_diagnostics}
 \resizebox{\textwidth}{!}{%
 \begin{tabular}{lccc}
  \hline\hline
  Curvature prior & Acceptance fractions & Split-$\widehat R$ &
  Approximate $N_{\rm eff}$ \\
  \hline
  $\lambda\in(-0.03,0.03)$
  & $(0.1583,0.1564,0.1565,0.1572)$
  & $(1.00017,1.00031,1.00021,1.00019)$
  & $(9806,8838,9176,10940)$ \\
  $\lambda\in(-0.01,0.01)$
  & $(0.1760,0.1787,0.1748,0.1729)$
  & $(1.00137,1.00163,1.00146,1.00111)$
  & $(8021,7392,7531,9042)$ \\
  \hline\hline
 \end{tabular}%
 }
\end{table*}

\subsection{Posterior constraints and Kerr validation}

The marginalized estimates are given in Table~\ref{tab:qpo_posteriors}.  For
the primary prior, the 16th, 50th, and 84th percentiles yield
\begin{align}
 \frac{M}{M_\odot}
 &=5.423^{+0.296}_{-0.299},
 &
 a_\ast
 &=0.2831^{+0.0085}_{-0.0072},
 \nonumber\\
 x
 &=5.609^{+0.181}_{-0.162},
 &
 \lambda
 &=-5.57^{+12.94}_{-16.15}\times10^{-4}.
 \label{eq:qpo_primary_result}
\end{align}
The central credible intervals for the curvature are
\begin{align}
 -2.171\times10^{-3}<\lambda&<7.373\times10^{-4}
 &&(68\%),\nonumber\\
 -4.035\times10^{-3}<\lambda&<1.731\times10^{-3}
 &&(95\%).
 \label{eq:qpo_lambda_intervals}
\end{align}
In particular, the Kerr value $\lambda=0$ lies within the $68\%$ interval.

\begin{table*}[t]
 \centering
 \caption{Posterior medians and central credible intervals for the two
 curvature priors. The close agreement of the two rows in every parameter
 demonstrates that the inference is insensitive to the tested prior width.}
 \label{tab:qpo_posteriors}
 \begin{tabular}{lcccc}
  \hline\hline
  Prior on $\lambda$
  & $M/M_\odot$
  & $a_\ast$
  & $x=r/M$
  & $\lambda=R_0M^2$ \\
  \hline
  $\mathcal{U}(-0.03,0.03)$
  & $5.423^{+0.296}_{-0.299}$
  & $0.2831^{+0.0085}_{-0.0072}$
  & $5.609^{+0.181}_{-0.162}$
  & $-5.57^{+12.94}_{-16.15}\times10^{-4}$ \\
  $\mathcal{U}(-0.01,0.01)$
  & $5.423^{+0.296}_{-0.299}$
  & $0.2831^{+0.0085}_{-0.0072}$
  & $5.609^{+0.181}_{-0.162}$
  & $-5.68^{+12.97}_{-16.16}\times10^{-4}$ \\
  \hline\hline
  \multicolumn{5}{l}{\footnotesize The quoted uncertainties delimit the
  central $68\%$ credible intervals.}
 \end{tabular}
\end{table*}

The maximum-likelihood point of the primary run is
\begin{equation}
 \left(M/M_\odot,a_\ast,x,\lambda\right)_{\rm ML}
 =
 \left(
 5.400,\,
 0.28349,\,
 5.6217,\,
 -4.519\times10^{-4}
 \right),
 \label{eq:qpo_best_fit}
\end{equation}
with
\begin{equation}
 \chi^2_{\min}<10^{-9}.
 \label{eq:qpo_minimum_chi2}
\end{equation}
The corresponding model frequencies are
\begin{equation}
 \left(
 \nu_U^{\rm th},
 \nu_L^{\rm th},
 \nu_C^{\rm th}
 \right)
 =
 \left(
 441.000,\,
 298.000,\,
 17.300
 \right)\,{\rm Hz}.
 \label{eq:qpo_best_frequencies}
\end{equation}
Thus, the predicted frequencies agree with the observed QPO triplet
to substantially better precision than the observational
uncertainties. The narrow-prior calculation yields the same optimum
to the quoted precision and also gives
$\chi^2_{\min}<10^{-9}$. The extremely small residuals are numerical
consequences of a saturated inference problem: the three QPO
frequencies and the dynamical-mass term provide four constraints for
four fitted parameters. They verify the numerical solution of the
frequency equations, but they do not constitute an independent
goodness-of-fit test of the RP identification.

As a separate consistency check, we impose $\lambda=0$ and optimize
the three remaining parameters. This gives
\begin{equation}
 \left(
 \frac{M}{M_\odot},a_\ast,x
 \right)_{\rm Kerr}
 =
 \left(
 5.3044,\,
 0.28598,\,
 5.6780
 \right),
 \label{eq:qpo_kerr_validation}
\end{equation}
and $\chi^2_{\rm Kerr}=0.1061$, reproducing the established Kerr RP
result $M\simeq5.31M_\odot$ and $a_\ast\simeq0.29$ \cite{Motta:2013wga}. The profile
improvement of the four-parameter solution relative to Kerr is
therefore only
\begin{equation}
 \Delta\chi^2(\lambda=0)=0.1061.
 \label{eq:qpo_delta_chi2}
\end{equation}
For one additional parameter, such a small change provides no
statistical motivation for departing from Kerr. The agreement with the established Kerr RP solution checks the
$R_0=0$ implementation and the conversion to physical frequency
units. Since $\Xi=1$ in this limit, this comparison does not
independently validate the $\Xi$ dependence at nonzero curvature
or the identification of the stationary time with the
observational clock.

\subsection{Posterior geometry and prior sensitivity}

\begin{figure*}[!t]
 \centering
 \includegraphics[width=0.92\textwidth]{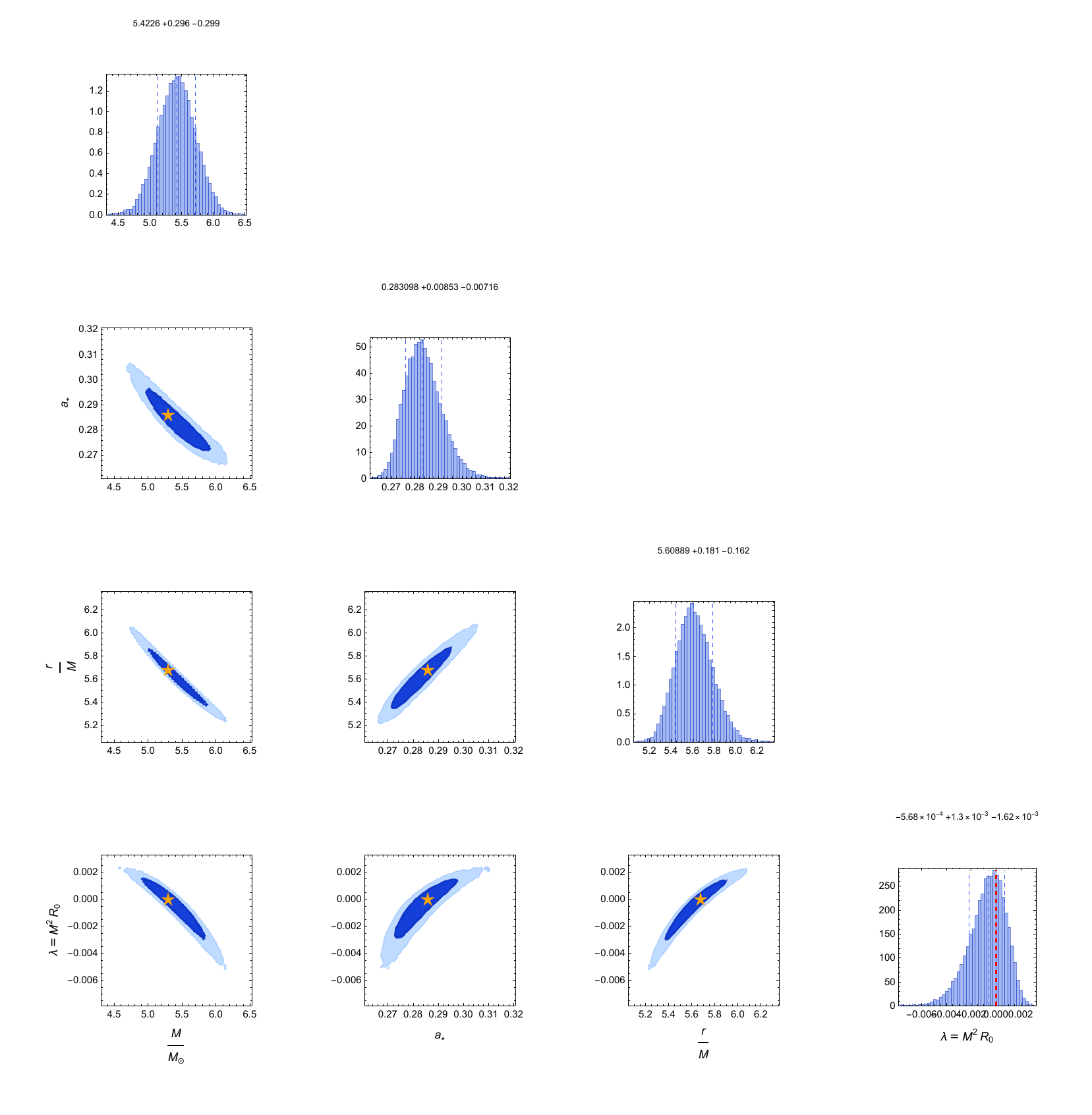}
 \caption{Marginalized posterior distributions for GRO J1655--40
obtained in the prior-sensitivity run,
$\lambda\in(-0.01,0.01)$. The diagonal dashed lines denote the 16th,
50th, and 84th percentiles. From dark to light blue, the
off-diagonal regions show the conventional two-dimensional
$1\sigma$, $2\sigma$, and $3\sigma$ HPD contours, approximately
enclosing $39.3\%$, $86.5\%$, and $98.9\%$ of the joint posterior
probability. The gold stars mark projections of the Kerr best-fit
point, and the red line in the curvature histogram denotes
$\lambda=0$.}
 \label{fig:qpo_corner_narrow}
\end{figure*}

\begin{figure*}[!t]
 \centering
 \includegraphics[width=0.92\textwidth]{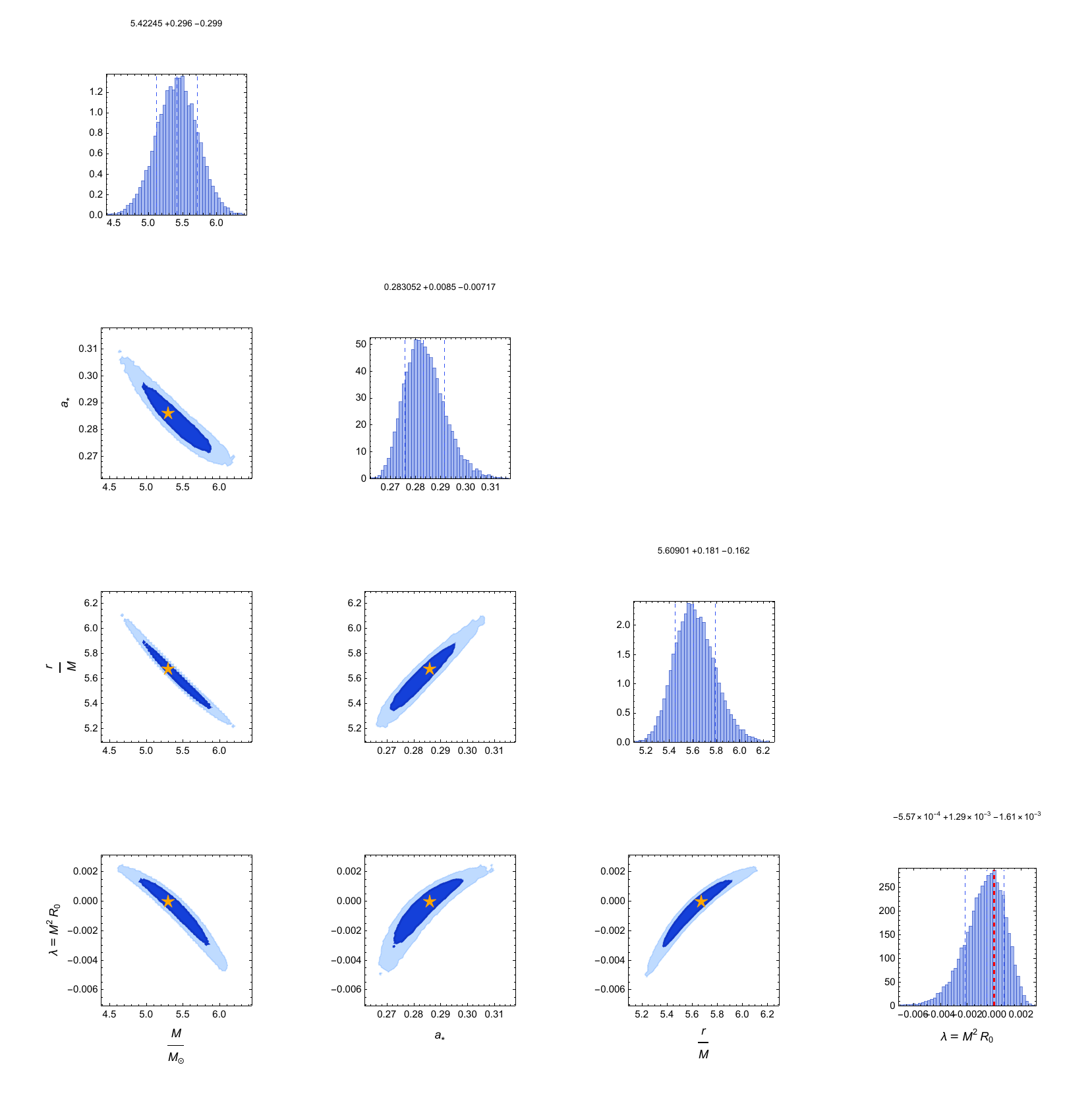}
 \caption{Marginalized posterior distributions for the primary inference,
 $\lambda\in(-0.03,0.03)$.  The percentile lines, HPD probabilities, Kerr
 markers, and color convention are identical to those in
 Fig.~\ref{fig:qpo_corner_narrow}.  The near identity of the two figures shows
 that the sampled likelihood is already localized well inside the narrower
 curvature prior.}
 \label{fig:qpo_corner_primary}
\end{figure*}

Figures~\ref{fig:qpo_corner_narrow} and
\ref{fig:qpo_corner_primary} show the one- and two-dimensional
marginalized posteriors for the narrow and primary priors,
respectively. The diagonal panels display the 16th, 50th, and 84th
percentiles, whereas the off-diagonal highest-posterior-density
contours, estimated from binned and smoothed posterior densities,
approximately enclose $39.3\%$, $86.5\%$, and $98.9\%$ of the
two-dimensional joint posterior probability, corresponding to the
conventional Gaussian $1\sigma$, $2\sigma$, and $3\sigma$ regions.
For visualization only, the plotting ranges are based on the
percentiles corresponding to cumulative probabilities of $0.1\%$ and $99.9\%$ for each marginalized distribution, with
the Kerr reference point retained within the displayed ranges. Thus,
only the extreme $0.1\%$ tail at each end is omitted from the plotted
ranges, whereas all quoted posterior quantiles and credible intervals
are calculated from the complete retained samples.

The two figures exhibit the same elongated degeneracy directions.  The mass is
anticorrelated with $a_\ast$, $x$, and $\lambda$, while the latter three
parameters are mutually positively correlated.  These correlations follow
from the structure of Eq.~\eqref{eq:rp_identification}: changes in the overall
mass scale can be compensated by correlated changes in the spin, emission
radius, and curvature while maintaining the three measured frequencies.  The
particularly narrow directions in the $M$--$x$ and $x$--$\lambda$ planes show
that the data constrain combinations of parameters more strongly than any one
parameter in isolation.

The numerical differences between the two runs are negligible.  The medians
shift by only $1.5\times10^{-4}$ in $M/M_\odot$,
$4.6\times10^{-5}$ in $a_\ast$, $1.2\times10^{-4}$ in $x$, and
$1.1\times10^{-5}$ in $\lambda$.  Their $68\%$ and $95\%$ curvature
intervals also agree within the expected finite-chain fluctuations.  Since the
primary $95\%$ posterior support is approximately
$-4.0\times10^{-3}<\lambda<1.7\times10^{-3}$, it lies comfortably inside
both tested prior ranges.  The similarity of the two figures is therefore an
expected and desirable result: it demonstrates posterior contraction relative
to the prior and excludes prior-edge truncation as the origin of the reported
bound.
\clearpage

\subsection{Dependence on the dynamical-mass likelihood}
\label{sec:mass-likelihood-sensitivity}

The baseline curvature constraint is obtained from the joint
QPO-plus-dynamical-mass posterior.  Since the three QPO frequencies
constrain four model parameters, $(M/M_{\odot},a_{\ast},x,\lambda)$,
the timing data alone do not independently determine all four
quantities.  We therefore performed two additional controlled
calculations to quantify the role of the external mass information.
First, we doubled the standard deviation of the dynamical-mass
likelihood and adopted $M_{\rm dyn}=5.4\pm0.6\,M_{\odot}$.  Second, we
removed the Gaussian dynamical-mass term and retained only the three
QPO contributions to the likelihood.  In both calculations we used
the primary curvature prior $-0.03<\lambda<0.03$, the same uniform
priors on $M/M_{\odot}$, $a_{\ast}$, and $x$, and the same physical-domain
conditions and sampling settings as in the baseline analysis.

The additional chains satisfy the adopted convergence criterion.  For
the broadened-mass run, the acceptance fractions lie between $0.0774$
and $0.0815$, the largest split-$\widehat{R}$ is $1.00080$, and the
approximate effective sample sizes range from $1873$ to $2860$.  For
the QPO-only run, the acceptance fractions lie between $0.0287$ and
$0.0308$, the largest split-$\widehat{R}$ is $1.00100$, and the
effective sample sizes range from $1543$ to $1627$.  The acceptance
fractions are lower than in the baseline calculation, particularly for
the QPO-only posterior, because removing the mass likelihood produces
a narrow, strongly curved degeneracy.  Nevertheless, the agreement
among the four independent chains and the effective sample sizes are
sufficient for the diagnostic comparison presented here.

With the broadened mass uncertainty, the marginalized parameters are

\begin{align}
 \frac{M}{M_{\odot}}
 &=5.506^{+0.595}_{-0.578},
 &
 a_{\ast}
 &=0.2811^{+0.0166}_{-0.0116},
 \\
 x
 &=5.561^{+0.357}_{-0.300},
 &
 \lambda
 &=-9.80^{+23.80}_{-37.15}\times10^{-4},
\end{align}
where the quoted uncertainties delimit the central $68\%$ credible
intervals.  The corresponding curvature intervals are

\begin{align}
 -4.695\times10^{-3}
 &<\lambda<1.400\times10^{-3}
 &&(68\%),
 \\
 -9.425\times10^{-3}
 &<\lambda<2.632\times10^{-3}
 &&(95\%).
\end{align}

Thus, doubling the uncertainty of the external mass measurement
increases the widths of both the $68\%$ and $95\%$ curvature intervals
by factors of approximately $2.10$ relative to the baseline analysis.
The Kerr value $\lambda=0$ remains inside the $68\%$ interval. The best-fitting extended model remains near
\begin{equation}
 \boldsymbol{\vartheta}_{\rm ML}
 =
 \left(
 5.400,\,
 0.28349,\,
 5.6217,\,
 -4.519\times10^{-4}
 \right),
 \label{eq:qpo_broad_mass_best_fit}
\end{equation}
with $\chi^2_{\min}<10^{-9}$. Fixing $\lambda=0$ gives
$\chi^2_{\rm Kerr}=0.0274$, so allowing curvature to vary again
provides only a negligible profile-likelihood improvement.

For the QPO-only calculation, the formal marginalized estimates are

\begin{align}
 \frac{M}{M_{\odot}}
 &=5.779^{+0.837}_{-1.141},
 &
 a_{\ast}
 &=0.2754^{+0.0329}_{-0.0118},
 \\
 x
 &=5.418^{+0.709}_{-0.376},
 &
 \lambda
 &=-2.491^{+4.668}_{-6.622}\times10^{-3}.
\end{align}

The associated curvature intervals are

\begin{align}
 -9.113\times10^{-3}
 &<\lambda<2.178\times10^{-3}
 &&(68\%),
 \\
 -1.252\times10^{-2}
 &<\lambda<3.047\times10^{-3}
 &&(95\%).
\end{align}

These QPO-only intervals must not be interpreted as an independent
four-parameter measurement.  The joint posterior follows an extended
curved ridge, and the marginalized mass distribution approaches the
adopted uniform boundaries $4<M/M_{\odot}<7$.  Consequently, the
finite QPO-only interval for $\lambda$ is conditional on these mass
bounds and on the imposed physical-domain cuts.  Its $68\%$ width is
approximately $3.88$ times the baseline width.  The QPO-only Kerr and
extended models both reproduce the triplet with effectively vanishing
$\chi^2$, as expected for an underdetermined frequency-only problem.

The results are summarized in
Table~\ref{tab:mass-likelihood-sensitivity}. They demonstrate that
the external dynamical-mass measurement plays an essential role in
breaking the remaining QPO degeneracy. The primary curvature result
should therefore be described as a joint QPO and dynamical-mass
constraint, rather than as a constraint obtained from the QPO
triplet alone.

\begin{table*}[t]
\caption{Dependence of the inferred curvature on the dynamical-mass
likelihood.  All calculations use $-0.03<\lambda<0.03$ and the same
QPO measurements, uniform parameter bounds, physical-domain cuts, and
sampling settings.  The QPO-only intervals are conditional on the
uniform mass prior $4<M/M_{\odot}<7$ and should be regarded as a
prior-dependent diagnostic rather than as an independent measurement.}
\label{tab:mass-likelihood-sensitivity}
\centering
\begin{ruledtabular}
\begin{tabular}{lccc}
Mass likelihood
& $M/M_{\odot}$ $(68\%)$
& $10^{3}\lambda$ $(68\%)$
& $10^{3}\lambda$ $(95\%)$
\\
\hline
$5.4\pm0.3\,M_{\odot}$
& $5.423^{+0.296}_{-0.299}$
& $-0.557^{+1.294}_{-1.615}$
& $(-4.035,\,1.731)$
\\
$5.4\pm0.6\,M_{\odot}$
& $5.506^{+0.595}_{-0.578}$
& $-0.980^{+2.380}_{-3.715}$
& $(-9.425,\,2.632)$
\\
None (QPO only)
& $5.779^{+0.837}_{-1.141}$
& $-2.491^{+4.668}_{-6.622}$
& $(-12.516,\,3.047)$
\end{tabular}
\end{ruledtabular}
\end{table*}

\subsection{Observational interpretation and limitations}

The present inference should be viewed in the context of previous
QPO analyses of geometries containing a cosmological-constant
parameter
\cite{Rink2021QPO,Hesamolhokama2024QPO}.
In particular, Bayesian constraints on standard Kerr--de Sitter
spacetime from the GRO J1655--40 triplet have already been
considered~\cite{Hesamolhokama2024QPO}. Our analysis supplements
this line of investigation with controlled changes to the
curvature-prior width and dynamical-mass likelihood, together
with explicit verification of the exterior horizon domain.
Numerical comparisons between the inferred intervals require
consistent curvature definitions, frequency conventions,
likelihoods, and parameter priors.

The observational result should be interpreted as a model-dependent
strong-field constraint obtained from the joint
QPO-plus-dynamical-mass posterior. Within the geodesic RP
identification and the stationary-time normalization specified above,
the simultaneous QPO triplet of GRO J1655--40, when combined with the
independent dynamical-mass measurement, confines the dimensionless
curvature to the interval in
Eq.~\eqref{eq:qpo_lambda_intervals}. The agreement of the independent
chains, the curvature-prior comparison, the dynamical-mass-likelihood
sensitivity tests, and the exact recovery of the Kerr RP solution
demonstrate the numerical consistency of the inference. A sample-by-sample calculation of the horizon roots further confirms
that every retained QPO orbit lies outside the event horizon and, for
$\lambda>0$, inside the cosmological horizon.
These checks establish numerical consistency within the specified
forward model. They do not determine the physical relation between
the stationary coordinate time and the observer's clock.
In particular, the reported credible intervals do not include
systematic uncertainty associated with a different clock or
rotational-frame prescription, or with matching the effective
constant-curvature region to an exterior geometry.
The curvature constraint should therefore be interpreted as
conditional on the adopted frequency-to-observation mapping.

This robustness does not turn the result into evidence for nonzero curvature.
First, the marginal posterior contains $\lambda=0$ within $68\%$, and the
profile improvement over Kerr is only $\Delta\chi^2\simeq0.106$.  Second, the
RP assignments in Eq.~\eqref{eq:rp_identification} are phenomenological; disk
pressure, magnetic stresses, finite thickness, nongeodesic corrections, and a
radially extended emission region are not included.  Third, neutral geodesic
observables in the constant-curvature sector determine the effective geometry,
not the underlying function $f(R)$ uniquely.  The inferred parameter should
therefore be regarded as a bound on a local effective $R_0M^2$ within this
geometric model, rather than as a measurement of a particular $f(R)$
Lagrangian or of the cosmological curvature.

Accordingly, the statistically and physically defensible result is
\begin{equation}
 R_0M^2=-5.57^{+12.94}_{-16.15}\times10^{-4}
 \qquad (68\%~{\rm credible}),
\end{equation}
with the Kerr limit allowed. The combined QPO and dynamical-mass data
therefore provide a quantitative two-sided constraint on the
dimensionless curvature parameter $R_0M^2$, but no statistically
significant evidence for a departure from the Kerr geometry.
\clearpage

\section{Summary and discussion}
\label{sec:conclusions}
 
In this work, we have presented a unified analysis of equatorial circular-orbit stability, epicyclic resonances, relativistic precession, and QPO-based observational constraints in the constant-curvature vacuum sector of metric $f(R)$ gravity. For every nondegenerate branch satisfying the algebraic trace equation, the background field equations reduce to the Einstein-space condition~(\ref{effective-cosmological-constant}). Consequently, the neutral static and rotating geometries are described by the Schwarzschild--(anti-)de~Sitter and Kerr--(anti-)de~Sitter spacetimes, respectively, with $\Leff=R_0/4$. The underlying $f(R)$ model determines which values of $R_0$ are dynamically admissible and whether the corresponding branch is viable. Once a particular constant-curvature branch has been selected, however, the neutral geodesic observables considered here depend on the gravitational theory only through $R_0$.

We first collected the standard relations for equatorial circular
geodesics and their linear radial and vertical perturbations in stationary, axisymmetric, and reflection-symmetric spacetimes. The azimuthal, radial, and vertical frequencies were consistently defined with respect to the stationary coordinate time. This distinction is important because the individual coordinate-time frequencies depend on the normalization of the stationary time coordinate, whereas ratios of frequencies are invariant under a common constant rescaling. Radial and vertical stability require $\Omega_r^2>0$ and $\Omega_\theta^2>0$, respectively, while the conditions $\Omega_r^2=0$ and $\Omega_\theta^2=0$ determine the corresponding marginally stable orbits.

For the static Schwarzschild--(anti-)de~Sitter background, the fundamental frequencies reduce to Eqs.~\eqref{static-Omega-phi-theta} and~(\ref{static-Omega-r}). The equality $\Omega_\theta=\Omega_\phi$ follows exactly from spherical symmetry and implies that the nodal-precession frequency vanishes identically in the static geometry. The circular photon orbit remains located at $r=3M$ for every value of $R_0$, although the horizon structure, marginally stable orbits, and large-radius behavior of the orbital frequencies are modified by the background curvature.

The topology of the stable circular-orbit region depends qualitatively on the sign of $R_0$. For $R_0<0$, the radial-stability condition possesses a single physical root corresponding to an ISCO slightly inside the Schwarzschild value $r=6M$, and stable circular motion extends toward arbitrarily large radii. In the curvature-dominated regime,

\begin{equation*}
\Omega_r\longrightarrow\sqrt{\frac{|R_0|}{3}}, \qquad \Omega_\phi=\Omega_\theta\longrightarrow \sqrt{\frac{|R_0|}{12}},
\end{equation*}
so that $\Omega_\theta/\Omega_r\rightarrow1/2$. This asymptotic hierarchy differs qualitatively from the Schwarzschild limit, for which $\Omega_\theta/\Omega_r\rightarrow1$. For $R_0>0$, stable circular motion is instead confined to a finite annulus,

\begin{equation*}
\risco<r<\rosco.
\end{equation*}

The ISCO moves outward as $R_0$ increases, whereas the OSCO moves inward. The two marginally stable orbits merge at

\begin{equation*}
\risco=\rosco=\frac{15M}{2},
\end{equation*}
when the curvature reaches

\begin{equation*}
R_0^{\rm crit}M^2=\frac{16}{5625} \simeq2.84\times10^{-3}.
\end{equation*}
Above this value, no radially stable timelike circular orbit exists in the static geometry.

A direct consequence of the ISCO--OSCO confinement is the emergence of multiple resonance branches. For $R_0>0$, the radial epicyclic frequency vanishes at both boundaries of the stable annulus, and therefore $\Omega_\theta/\Omega_r\rightarrow+\infty$ near both the ISCO and the OSCO. The frequency ratio develops a minimum between these boundaries, allowing each of the resonance conditions

\begin{equation*}
\frac{\Omega_\theta}{\Omega_r} =\frac{3}{2},\quad 2,\quad 3,
\end{equation*}
to be realized at two distinct stable radii whenever the corresponding rational value lies above the minimum of the frequency-ratio profile. These two solutions form an inner, ISCO-side branch and an outer, OSCO-side branch. As $R_0$ increases, the two branches approach one another, merge, and eventually disappear before the entire stable annulus closes. By contrast, for $R_0\leq0$, each resonance condition is realized at a single stable radius.

In the rotating Kerr--(anti-)de~Sitter background, frame dragging removes the degeneracy between the azimuthal and vertical frequencies. The resulting expressions reproduce the static formulas in the limit $a\rightarrow0$ and the standard Kerr frequencies when $R_0\rightarrow0$. Increasing the corotating spin shifts the ISCO and the associated resonance radii inward and enlarges the curvature interval over which the outer resonance branches survive. The signed nodal-precession frequency $\Onod$, defined in Eq.~\eqref{nodal-definition}, is given to first order in the spin by Eq.~\eqref{nodal-linear}. Its first term is the standard Lense--Thirring contribution, while the second is the leading constant-curvature correction in the adopted coordinate-time normalization. For $R_0<0$, this correction opposes the usual frame-dragging contribution and reverses the sign of the nodal precession beyond the characteristic radius given approximately by Eq.~\eqref{nodal-reversal}. Sufficiently extended tilted flows may therefore contain inner and outer regions subject to precessional torques of opposite signs.

Using an angular-momentum-weighted rigid-flow model, we showed that the local curvature corrections propagate to the global precession frequency. For $R_0>0$, the OSCO provides a natural upper bound,

\begin{equation*}
r_{\mathrm{out}}\leq r_{\mathrm{OSCO}},
\end{equation*}
on the radial extent of the coherently precessing region. For $R_0<0$, the local reversal of $\Omega_{\mathrm{nod}}$ can produce a global reversal at a larger outer radius once the retrograde contribution from the outer flow compensates the prograde torque generated by the inner region. The corresponding alignment-time estimate inherits the curvature dependence through the scaling

\begin{equation*}
t_{\mathrm{align}}\propto\Omega_p^{-2}.
\end{equation*}

Its formal divergence at $\Omega_p=0$ indicates the breakdown of the simplified rigid-precession prescription near the torque-cancellation point rather than a physically infinite alignment time.

We subsequently confronted the rotating constant-curvature geometry with the simultaneous QPO triplet of GRO J1655--40 within the relativistic-precession model. Identifying the observed frequencies as

\begin{equation*}
\nu_U=\nu_\phi,\qquad \nu_L=\nu_\phi-\nu_r,\qquad \nu_C=\nu_\phi-\nu_\theta,
\end{equation*}

and combining them with the independent dynamical-mass measurement, the MCMC analysis yielded

\begin{equation*}
 R_0M^2 = -5.57^{+12.94}_{-16.15}\times10^{-4}  \qquad (68\%~{\rm credible}).
\end{equation*}

The corresponding central credible intervals are

\begin{equation*}
-2.171\times10^{-3} < R_0M^2 < 7.373\times10^{-4} \qquad (68\%),
\end{equation*}

and

\begin{equation*}
-4.035\times10^{-3} < R_0M^2 < 1.731\times10^{-3} \qquad (95\%).
\end{equation*}

The Kerr value $R_0M^2=0$ therefore lies within the $68\%$ credible interval. Moreover, fixing $R_0M^2=0$ reproduces the established Kerr relativistic-precession solution,

\begin{equation*}
\frac{M}{M_\odot}=5.3044,\qquad
a_\ast=0.28598,\qquad
\frac{r}{M}=5.6780,
\end{equation*}
and the profile-likelihood difference between the Kerr solution and the extended-model optimum is only

\begin{equation*}
\Delta\chi^2(\lambda=0)\simeq0.1061.
\end{equation*}

Allowing the curvature parameter to vary therefore produces no statistically meaningful improvement over Kerr.

The observational inference is numerically stable. Repeating the analysis with the narrower uniform prior $R_0M^2\in(-0.01,0.01)$ produces posterior medians and credible intervals nearly identical to those obtained with the primary prior $R_0M^2\in(-0.03,0.03)$. The convergence diagnostics also satisfy $\widehat R<1.01$, with effective sample sizes of several thousand for all inferred parameters. A sample-by-sample calculation of the
horizon roots also confirms that all retained QPO orbits lie outside
the event horizon and, for $R_0M^2>0$, inside the cosmological
horizon. The similarity of the two posterior distributions demonstrates that the curvature constraint is not produced by truncation at the tested prior boundaries. Nevertheless, the nearly vanishing minimum $\chi^2$ should not be interpreted as an independent goodness-of-fit test, because the three QPO frequencies and the dynamical-mass term provide four constraints for the four fitted parameters. The robust observational conclusion is therefore a model-dependent constraint on $R_0M^2$, not a detection of nonzero curvature.

The inferred curvature interval depends appreciably on the external
dynamical-mass information. Doubling the mass uncertainty from
$0.3M_{\odot}$ to $0.6M_{\odot}$ increases the widths of the
$68\%$ and $95\%$ curvature intervals by approximately a factor of
$2.1$, while the Kerr value remains allowed. When the Gaussian mass
likelihood is removed, the posterior develops an extended degeneracy
that approaches the adopted uniform mass boundaries. The resulting
finite curvature interval is therefore conditional on the parameter
priors and physical-domain cuts. Accordingly, the primary result must be interpreted as a joint QPO
and dynamical-mass constraint rather than as an independent QPO-only
measurement. The numerical interval is also conditional on the adopted
identification of the stationary coordinate-time frequencies
with the observed QPO frequencies. Its quoted uncertainty does
not include systematic uncertainty associated with an alternative
observer prescription or a global matching of the effective
constant-curvature region to the source environment.

The interpretation of the complete analysis requires three qualifications. First, the effects derived here are geometric consequences of $R_0$ and are not unique signatures of metric $f(R)$ gravity. All neutral geodesic results apply equally to Einstein gravity with $\Leff=R_0/4$. Observations based solely on neutral circular geodesics may therefore constrain the effective curvature, but they cannot determine the functional form of $f(R)$. Breaking this degeneracy would require observables sensitive to the additional scalar degree of freedom, such as gravitational perturbations, quasinormal modes, or matter configurations with nonconstant curvature.

Second, the values $|R_0|M^2\sim10^{-6}$--$10^{-3}$ explored here must be regarded as a phenomenological strong-field range. The observed cosmological constant corresponds to a dimensionless curvature far too small to affect black-hole orbital frequencies appreciably. Whether a viable metric $f(R)$ model can support an approximately constant local-curvature exterior with the magnitude constrained here is a separate, model-dependent question.

Third, although Sec.~\ref{sec:qpo_constraints} provides a quantitative geodesic-level comparison with the simultaneous QPO triplet of GRO J1655--40, interpreting the resulting curvature constraint as a precision test of the underlying gravitational theory would require a self-consistent accretion-flow model incorporating mode excitation, pressure, magnetic stresses, finite disk thickness, warp propagation, dissipation, and spectral modulation. The relativistic-precession identification is phenomenological, and the present analysis does not incorporate possible nongeodesic shifts of the characteristic frequencies or a radially extended emission region.

Within these limitations, the analysis identifies three robust
geometric consequences of a constant background curvature: a compact
ISCO--OSCO stability domain and doubled resonance branches for
$R_0>0$; the universal asymptotic relation
$\Omega_\theta/\Omega_r\rightarrow1/2$ for $R_0<0$; and, in rotating
backgrounds, a possible reversal of the signed nodal-precession frequency
relative to the adopted azimuthal frame. In addition, the GRO J1655--40 analysis establishes a
quantitative constraint on the effective dimensionless curvature
that is stable against the tested curvature-prior variation but
depends appreciably on the external dynamical-mass information. The
result remains fully consistent with the Kerr limit. These findings
provide a coherent foundation for future studies incorporating
realistic accretion dynamics and perturbative observables capable of
probing the gravitational theory beyond its constant-curvature
background geometry.

\begin{acknowledgments}
The work of KN is supported financially by the INSF of Iran under grant number 40501739. MB is supported by Proyecto Interno UCM-IN-25202 l\'inea regular, and FONDECYT grant 1262452. In preparing this manuscript, the authors used ChatGPT (OpenAI, GPT-6 Astra)
to assist with the development and debugging of Mathematica code
for data analysis and Markov chain Monte Carlo (MCMC) sampling,
as well as to improve language quality, grammar, and \LaTeX{}
typesetting. AI-generated code and textual suggestions were
critically reviewed and, where necessary, modified by the authors.
The authors executed the numerical calculations and checked the
resulting outputs, including the Kerr-limit validation, convergence
diagnostics, posterior constraints, and physical-domain conditions.
The authors retain full responsibility for the originality,
accuracy, and integrity of the intellectual content, including
all code, analyses, interpretations, and conclusions.

\end{acknowledgments}

\section*{Conflict of Interest}
The authors declare no conflict of interest.

\section*{Data Availability Statement}
The observational data used in this study are taken from the
published sources cited in the text. The Mathematica code and
posterior samples supporting the numerical inference are provided
in the accompanying supplementary material.


\end{document}